 \documentclass[final,3p,times,authoryear]{elsarticle}

\usepackage{amssymb}
\usepackage{amsthm}

\usepackage{amsmath}	
\usepackage{mathptmx}
\usepackage{txfonts}
\usepackage{hyperref}
\usepackage{cite}
\usepackage{float}
\usepackage{array}
\usepackage{tablefootnote}
\usepackage{tabu}
\usepackage{longtable}
\usepackage{graphicx}
\usepackage[export]{adjustbox}
\usepackage{cleveref}
\usepackage{color}
\definecolor{pink}{rgb}{1.0, 0.4, 0.6}
\definecolor{green}{rgb}{0.0, 0.5, 0.0}

\usepackage{titlesec}

\titleformat{\paragraph}
{\normalfont\normalsize\itshape}{\theparagraph}{1em}{}
\titlespacing*{\paragraph}
{0pt}{3.25ex plus 1ex minus .2ex}{0ex plus .2ex}
\usepackage[normalem]{ulem}

\journal{Icarus}

\begin{document}

\begin{frontmatter}



\title{Rings of non-spherical, axisymmetric bodies}


\author[adrs1,adrs4]{Akash Gupta\corref{cor1}}
\ead{akashgpt.iitk@gmail.com}
\author[adrs2]{Sharvari Nadkarni-Ghosh}
\author[adrs3,adrs4]{Ishan Sharma}

\address[adrs1]{Department of Aerospace Engineering, IIT Kanpur, Kanpur, UP, 208016, India}
\address[adrs2]{Department of Physics, IIT Kanpur, Kanpur, UP, 208016, India}
\address[adrs3]{Department of Mechanical Engineering, IIT Kanpur, Kanpur, UP, 208016, India}
\address[adrs4]{Mechanics \& Applied Mathematics Group, IIT Kanpur, Kanpur, UP, 208016, India}

\begin{abstract}
We investigate the dynamical behavior of rings around bodies whose shapes depart considerably from that of a sphere. To this end, we have developed a new self-gravitating discrete element N-body code, and employed a local simulation method to simulate a patch of the ring. The central body is modeled as a symmetric (oblate or prolate) ellipsoid, or defined through the characteristic frequencies (circular, vertical, epicyclic) that represent its gravitational field. Through our simulations we explore how a ring's behavior -- characterized by dynamical properties like impact frequency, granular temperature, number density, vertical thickness and radial width -- varies with the changing gravitational potential of the central body. We also contrast  properties of rings about large central bodies (e.g. Saturn) with those of smaller ones (e.g. Chariklo). Finally, we investigate how the characteristic frequencies of a central body, restricted to being a solid of revolution with an equatorial plane of symmetry, affect the ring dynamics. The latter process may be employed to qualitatively understand the dynamics of rings about any symmetric solids of revolution.
\end{abstract}


\end{frontmatter}


\section{Introduction}

\par Rings surrounding the gas and ice giants are one of the more fascinating sights in our solar system. These entities display subtle dynamics that lead to intricate structures, not all of which are explained. Rings also act as unique mini-laboratories, involving smaller time-scales, that could help in advancing our understanding of astrophysical disks.

\par Recently, \citet{braga2014a} discovered two narrow rings about a non-spherical small-body Chariklo (equatorial radius $\sim 144.9$ km and oblateness $\sim 0.213$), a centaur orbiting the Sun. There is also speculation about a ring system of another minor-planet, Chiron \citep{ortiz2015a}, and it is suspected that there might have been rings around Saturn's moon Rhea, which is triaxial in shape \citep{jones2008a, tiscareno2010a}, or around Iapetus, which is oblate \citep{ip2006a}. Moreover, hot `Jupiter' exo-planets are likely candidates as ring hosts and those exo-planets which are tidally locked with their parent stars may have non-axisymmetric tidal bulges. Thus, the occurrence of rings around non-spherical, irregular bodies may be a more general phenomenon than anticipated. This raises the question as to how such rings may come into existence, and whether such structures are dynamically stable or, are they transient in nature. The study of rings may also help in understanding the primary body's interior \citep{hedman2013a}.

\par There is much work on planetary rings, but not as much in the context of rings around non-spherical bodies like asteroids, satellites or exo-planets. Some, recent studies have investigated formation scenarios for rings around smaller bodies like Chariklo; see \citet{pan2016a}, \citet{hyodo2016a} and \citet{araujo2016a}. These studies hypothesize that rings around Chariklo, or similar bodies, could come into being through tidal disruption of the central body by a planet, disruption of a previous satellite, or out-gassing of material from the central body itself. Earlier, \citet{dobrovolskis1989a} analyzed the stability of rings in the host body's equatorial/symmetry plane, with the latter being an oblate or prolate object. They observed that, while this equatorial plane is an energy minimum for oblate bodies, it is an energy maximum for prolate bodies. However, according to them, if a debris disk does come into existence around a prolate body, it could successfully form into a ring owing to dissipative collisions, in the same manner as for a debris disk around an oblate body. In a related recent work, \citet{lehebel2015a} study the motion of a ring particle around a body that has a non-axisymmetric bulge in the equatorial plane. They show that, under some conditions on the rotation period of the central body, the orbital period of the particle, and the particle's orbital precessional period, the central body's bulge can be averaged and incorporated as an effective $J_2$ correction.


\par One way to investigate the dynamics of rings is to model them as particulate systems and simulate them. Earlier works on planetary rings were concerned with simulating the complete ring system, e.g. works by \citet{trulsen1971a, trulsen1972a, trulsen1972b}, \citet{brahic1975a, brahic1977a} and \citet{hameen1978a, hameen1981a, hameen1982a}. These early simulations evolved the entire trajectory of all constituent ring particles; these particles were large in size and few in number. Though these simulations gave preliminary insights into the ring dynamics, they did not resolve finer dynamical details. Even today, the number of particles considered ranges between $10^3-10^6$, depending on model complexity and the corresponding computational costs. However, even the higher limit of this range is tiny in comparison to the particle distribution in a typical ring system. \citet{wisdom1988a} proposed a local cell simulation method in planetary ring simulations, which allowed the consideration of a much larger number of particles. Such simulations have allowed further insights into the ring dynamics. Recent studies, like \citet{salo1995a}, \citet{salo2010a}, \citet{lewis2001a}, \citet{lewis2000a}, \citet{perrine2011a} etc., have primarily followed this simulation methodology.

\par In this work, we numerically investigate the dynamics around non-spherical bodies, specifically symmetric ellipsoids that may be oblate or prolate. We also employ the local simulation method wherein the particles are restricted to a small patch and periodic boundary conditions are applied in the plane of the ring. In addition to the gravity of the central body, the particles in the patch are allowed to  interact through self-gravity and collisions. To account for collisions, we use the Discrete Element (DE) code of \citet{bhateja2016a}. This code is based on the spring-dashpot model proposed by \citet{cundall1979a} to model inelastic collisions between particles.

\par The paper is subsequently divided into four parts. In the next section, we discuss our simulation methodology. In Sec. 3 we briefly present details on how the code is implemented. Following this, we discuss the results of our study in Sec. 5 and 6. We conclude in Sec. 7 with a brief summary and thoughts about future work. Appendices present additional details of our simulation algorithm, code validation and the relevant parameters of study.



\section{Simulation methodology} \label{sim_method}
\subsection{Test-Section}
\par We follow the local simulation method and focus on a `Test-Section' (TS) with $N$ particles that is orbiting about a central/primary body; see \Cref{fig:ellipsoid_TS}. This central body is assumed to have a ring about it and, at any instant of time, the particles contained in the TS represent those in a small patch of the ring. The central body is such that it produces an axisymmetric gravitational field that is also symmetric about the body's equatorial plane. The equatorial plane is also the mid-plane of the ring.

\par The TS is a rectangular box whose mid-plane coincides with the central body's equatorial plane. The spatial position of the TS is defined by its centroid \textit{C}, also referred to as the `guiding center'. This point \textit{C} is situated at a fixed radial distance $r_{gc}$ from the center of the central body and moves on a circular orbit at the corresponding Keplerian angular velocity $\Omega$; this defines the orbiting velocity of the TS.

\par To describe the motion of particles inside the TS, a rotating Cartesian coordinate system $(x\;,\;y,\;z)$ is attached to \emph{C}; see \Cref{fig:ellipsoid_TS,fig:mnras4}. The \textit{x}-axis points in the radial direction away from the central body, the \textit{y}-axis is in the direction of the orbital motion of the TS, and the \textit{z}-axis is normal to the equatorial plane. The TS is oriented such that a pair of its faces are normal to \textit{x} and \textit{y} axes. The \textit{z}-direction is unbounded. The dimensions of the TS in \textit{x} and \textit{y} directions are $x_{TS}$ and $y_{TS}$, respectively. These dimensions are much smaller than the orbital radius $r_{gc}$. This justifies our ignoring the curvature of ring and, consequently, linearizing the equations governing particle motion in the TS. The dimensions of the TS must, however, be large enough to not allow any significant correlation in the dynamics of particles present at its opposite faces. 

\par A significant part of this study investigates the dynamics of rings about an axisymmetric ellipsoidal central body. Thus, in the next section, we discuss the gravitational potential of an axisymmetric ellipsoid. Following this, we describe the equations of motion of particles in the TS and the applicable boundary conditions.

\subsection{Gravitational potential of an axisymmetric ellipsoid}

\par
\begin{figure}[t]
\centering
\includegraphics[width=0.4\textwidth]{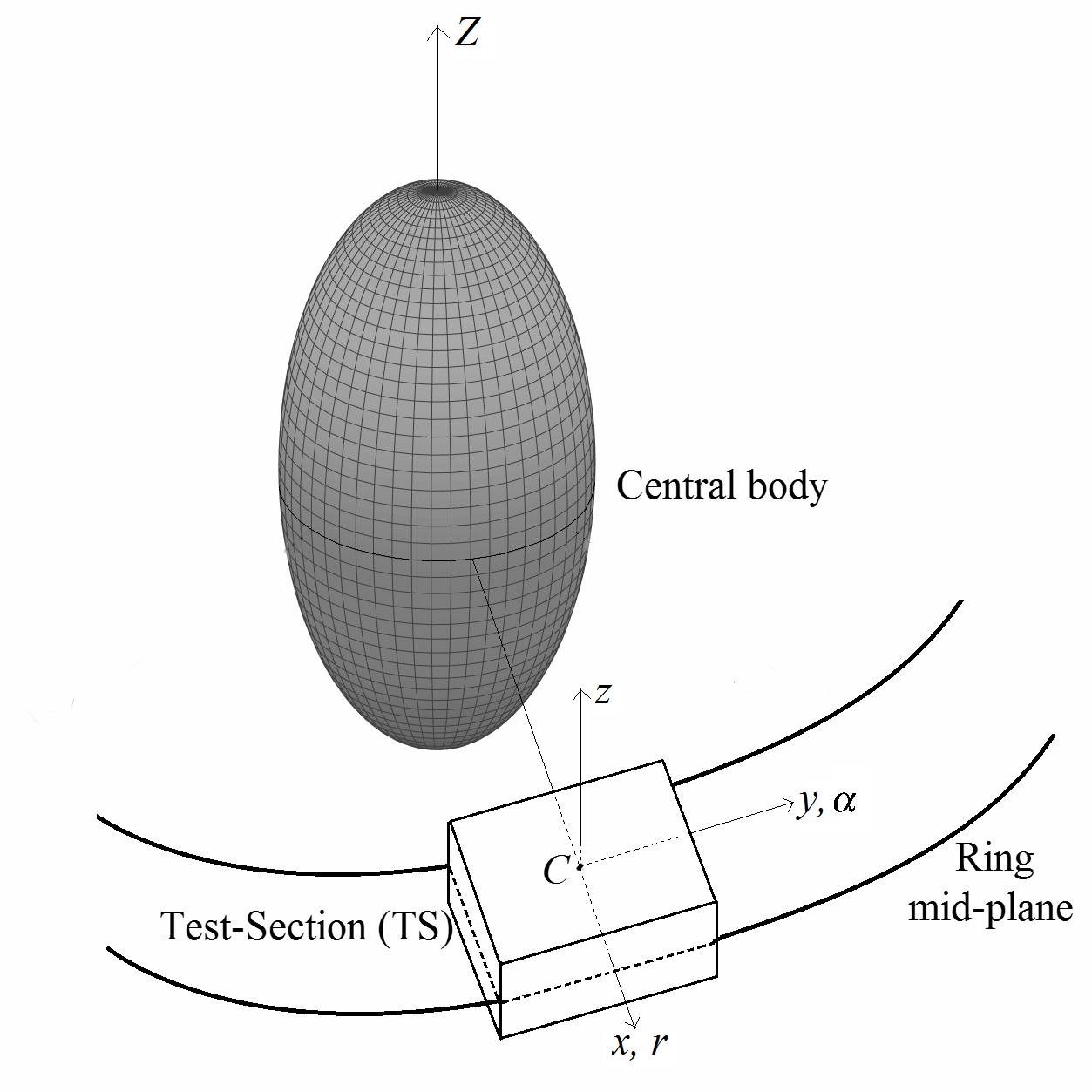} 
\caption{Schematic representation of the test-section (TS) orbiting about an ellipsoidal central body. Also shown is a cylindrical coordinate system $(r,\;\alpha,\;Z)$ with origin \textit{O} at the center of the central body, and a Cartesian coordinate system $(x,\;y,\;z)$ with origin \textit{C} (guiding center) at the center of the TS and which orbits with the TS. The axes \textit{Z} and \textit{z} are parallel.}
\label{fig:ellipsoid_TS}
\end{figure}

 The gravitational potential of an axisymmetric ellipsoid is computed in a cylindrical coordinate system $(r,\;\alpha,\;Z)$  with origin \textit{O} at the ellipsoid's center, and the \textit{Z}-axis along the axis of symmetry of the ellipsoid; see \Cref{fig:ellipsoid_TS}. The ellipsoid is characterized by its homogeneous density $\rho_{cb}$, and its semi-major axes, $a_r$ (in equatorial plane) and $a_z$ (along axis of symmetry). Simplifying the analytical formulation for a homogeneous triaxial ellipsoid (\citealp[Appendix B]{sharma2017a}; \citealp{sharma2010a}; \citealp[p. 38]{chandrashekhar1969a}), the gravitational potential of a homogeneous, axisymmetric, ellipsoidal body at an exterior point $(r,\;\alpha,\;Z)$ may be written as 

\begin{equation}\label{eq:ell_pot}
\Phi\left(r, Z\right) = \frac{\beta}{2} a_{r}^2 a_{Z} \int^\infty_{\lambda (r,Z)} \phi\left(r,Z;u \right)\,\frac{\textrm{d}u}{\Delta(u)},
\end{equation}
with \textit{G} being the universal gravitational constant, and
\begin{align}
\Delta (u) &= \sqrt{(a_{r}^2 + u)^2 (a_{Z}^2 + u)},\\
\phi \left( r,Z; u\right) &= \frac{r^2}{a_{r}^2 + u} +\frac{Z^2}{a_{z}^2 + u }-1, \\
\beta &= 2 \pi \rho_{cb} G\\
\textrm{and}\;\; \phi \left( r,Z; \lambda\right) &= 0,\label{eq:lambda}
\end{align}
where $\lambda$ is the ellipsoidal coordinate of $(r,\;\alpha,\;Z)$ defined by the ellipsoid passing through $(r,\;\alpha,\;Z)$ and confocal to the ellipsoidal central body. We obtain $\lambda$ as the largest root of (\ref{eq:lambda}). Because the potential field is axisymmetric, it has no $\alpha$ dependence.

\subsection{Equations of motion}\label{B}
\par We now describe the equations governing the motion of a ring particle in the TS, relative to a reference frame attached to the TS at \textit{C}; see \Cref{fig:ellipsoid_TS}. The effect of the central body's gravity will be linearized about \textit{C}. In the Cartesian coordinate system $(x,\;y,\;z)$ these equations may be expressed as \citep[p. 159]{binney2008a}
\begin{subequations}\label{eq:eqm123}
\begin{align}
\ddot{x} - 2\Omega \dot{y} \;\;+\;\; (\kappa^2 - 4\Omega^2)x \;\;&=\;\; \frac{F_x}{m}, \label{eq:eqm1}\\
\ddot{y} \;\;+ \;\;2\Omega \dot{x} \;\;&=\;\; \frac{F_y}{m} \label{eq:eqm2}\\ 
\textrm{and}\;\;\ddot{z}\;\; + \;\;\nu^2 z \;\;&= \;\;\frac{F_z}{m} \label{eq:eqm3},
\end{align}
\end{subequations}
where
\begin{subequations}\label{eq:eqm456}
\begin{align}
\kappa^2 &= \left[\frac{\partial^2 \Phi}{\partial r^2} + 3 \Omega^2 \right]_{(r=r_{gc},z=0)},\label{eq:eqm4}\\
\nu^2 &= \left[\frac{\partial^2 \Phi}{\partial z^2}\right]_{(r=r_{gc},z=0)}\label{eq:eqm5}\\
\textrm{and}\;\;\Omega^2 &= \left[ \frac{1}{r} \frac{\partial \Phi}{\partial r} \right]_{(r=r_{gc},z=0)},  \label{eq:eqm6}
\end{align}
\end{subequations}
are the squares of the epicyclic/radial, vertical and circular frequencies, respectively, $m$ is the mass of the ring particle, and $F_x$, $F_y$ and $F_z$ correspond to forces other than the central body's gravity, which are due to inter-particle gravitational attraction and dissipative collisions. When the central body is spherical, $\kappa = \nu = \Omega$ and (\ref{eq:eqm123}) reduces to the well-known Hill's equations \citep{hill1878a}. Note that the gravitational effect of the central body is only considered upto the second spatial derivative of its gravitational potential.

\begin{figure}[t]
\centering
\includegraphics[width=.5\columnwidth]{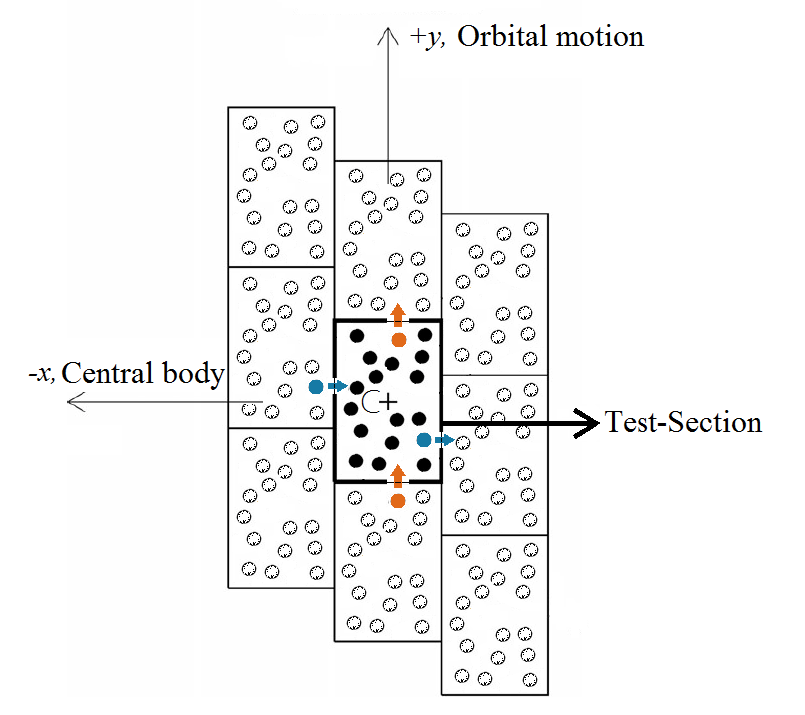}
\caption{Schematic representation of the test-section (TS) when viewed normal to plane of the ring along the $z$ direction. The TS is surrounded by identical images because of the imposed periodic boundary conditions; cf. \Cref{bound_cond}. Initially, boundaries of all the sections are aligned. However, they get misaligned with time due to Keplerian shear caused by the orbital motion of the TS; see \Cref{bound_cond} for more details. The figure also depicts particle movement across boundaries normal to the $x$- (blue) and $y$-(orange) axes, with the arrows indicating their direction of motion.}
\label{fig:mnras4}
\end{figure} 

\par The homogeneous solution to (\ref{eq:eqm123}) when $F_{x}=F_y=F_z = 0$ is
\begin{subequations}\label{eq:sol1}
\begin{align}
x &= x_o + P \cos(\kappa t) + Q \sin(\kappa t)\label{eq:sol1a},\\
y &= y_o + \Psi x_o t - 2P\frac{ \Omega}{\kappa} \sin(\kappa t) + 2Q\frac{ \Omega}{\kappa} \cos(\kappa t)\label{eq:sol1b}  \\
\textrm{and}\;\;z &= R \cos(\nu t) + S \sin(\nu t)\label{eq:sol1c},
\end{align}
\end{subequations}
where
\begin{equation}\label{eq:Psi}
\Psi = \frac{\kappa^2 - 4\Omega^2}{2 \Omega},
\end{equation}
and $x_o$, $y_o$, $P$, $Q$, $R$, and $S$ are constants of integration. The solution above may now be interpreted as having three components: mean motion about the central body with frequency $\Omega$, an epicyclic motion in the equatorial plane with frequency $\kappa$, and vertical oscillations normal to the equatorial plane with frequency $\nu$. The parameter $\Psi$ is related to the periodic boundary conditions and we discuss it in detail in the next subsection.

\subsubsection{Boundary conditions} \label{bound_cond}
\par A critical aspect of our simulations is the treatment of boundary conditions at the edges of the TS. The local simulation approach utilizes periodic boundary conditions. It is expected that the dynamics of particles in two regions of the ring separated by distances of the order of a few mean free paths should be statistically independent \citep{wisdom1988a}. Thus, it is assumed that the ring is comprised of numerous copies of a test-section (TS), as shown in \Cref{fig:mnras4}, and we are concerned with only one of these copies, namely TS. 
This, in turn, corresponds to imposing appropriate boundary conditions on TS.

\par If a particle crosses the boundary normal to the $y$-axis, it simply reappears at the opposite edge with the same velocity and at the same $x$-coordinate. However, boundary conditions along the $x$-axis are more complicated because of the Keplerian shear introduced by the orbital motion. Because all particles are assumed to be in nearly circular orbits, those close to the opposite boundaries normal to the $x$-axis will differ in their $y$-direction velocities by $\Psi x_{TS}$. Recall that $x_{TS}$ is the radial width of the TS, and $\Psi$ is the gradient of the $y$-direction (orbital) velocity, i.e., it corresponds to the Keplerian shear. This may be seen by differentiating (\ref{eq:sol1b}) with respect to time. Thus, particles exiting a boundary normal to the $x$-axis re-enter at the opposite face after having their $y$-velocity adjusted by $\pm \Psi x_{TS}$. At the same time, because radially adjoining test sections are separated by $x_{TS}$, they slide across each other with a relative velocity  $\Psi x_{TS}$. Recall that the orbital velocity of any test-section equals the orbital velocity of its `guiding center'. 
Because of this sliding, the relative placement of radially adjacent test-section images changes continually. Thus, when a particle exits a boundary normal to the $x$-axis, it may also experience a shift in its $y$-coordinate when reappearing at the opposite boundary. The extent of this shift will depend on the orbital velocity, $x_{TS}$ and the total time elapsed since the start of the simulation. This is illustrated in \Cref{fig:mnras4} that shows the test-sections at some time.
\par To summarize, when a particle moves across the boundary normal to $x$-axis, it also experiences a change in its $y$-velocity and, possibly, its $y$-coordinate, while when a particle goes across the boundary normal to $y$-axis, it simply reappears on the other-side with the same velocity and the same $x$ location.

\subsubsection{Constants of motion}
\par We now obtain `constants of motion' that help in checking the simulation's accuracy. These constants pertain to the center-of-mass velocities \citep{wisdom1988a}, and are
\begin{align*}
u = \frac{\sum_{i=1}^N m_i\dot{x_i}}{\sum_{i=1}^N m_i}\;\;\;&\textrm{and}\;\;\;w= \frac{\sum_{i=1}^N (\dot{y_i} - \Psi x_i)}{\sum_{i=1}^N m_i},
\end{align*}
where $m_i$ is the mass of individual ring particles. Note that the form of \emph{w} is different from the usual definition of center-of-mass velocity in the y-direction. This definition is chosen so that \textit{w} remains constant even if particles cross boundaries normal to the $x$-axis, which otherwise entails an adjustment in the $y$-velocity by $\Psi x_{TS}$. Our simulations limit fluctuations about the mean to $\sim10^{-16}$ ms$^{-1}$.

\section{Simulation algorithm}
\par We simulate particles in the ring through a soft-particle discrete element (DE) code that is developed on FORTRAN 90 and MATLAB. Important related issues are integration of the governing equations, implementation of boundary conditions, resolving collisional dynamics and accounting for self-gravity. Boundary conditions were discussed in \Cref{bound_cond}. For integration of governing equations we utilize a fourth-order Runge-Kutta scheme. To resolve collisions in DE simulations, we assume that particles in contact interact through a linear spring and a linear dashpot, whose deformations are modeled through `overlaps' between particles; cf. \Cref{fig:mnras1}. Inter-particle gravitational forces are approximated through mutual interaction between only those particles which lie within a certain cut-off distance (sphere of influence) of each other. A detailed discussion of these issues is available in \ref{sim_detail}. We have also validated our code against one employed in former studies and also through analytical calculations. These results are presented in \ref{code_val}.

\section{Simulation specifications} \label{sec:sim_spec}
\par
\begin{figure*}
\centering
\includegraphics[width=.6\textwidth]{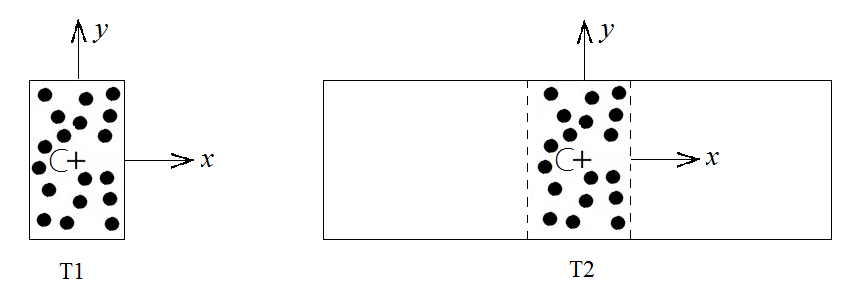}
\caption{Schematic representation of T1 and T2 cases (not to scale).}
\label{fig:mnras0_3}
\end{figure*}

 Our simulations are divided into two categories, namely, Type 1 (T1) and Type 2 (T2). In T1 simulations, the radial width of the ring is such that particles span the whole test-section, while in T2 simulations they initially only occupy a central region of radial width $300$ times the particle radius. The T2 case may be imagined as a localized region of a narrow ring, whereas T1 simulations correspond to a small patch of a much wider ring, as in this case the $x$-direction periodic boundary conditions come into play. \Cref{fig:mnras0_3} shows both T1 and T2 configurations. \Cref{table:TS_dim} reports the dimensions of the TS for each type of simulations. The choice of TS' dimensions ensure that its size does not influence ring's dynamical properties observed during simulations.

\begin{table}[h]
\centering
\begin{tabular}{ c  c  c } 
 \hline
 TS dimensions (in particle radii, $r_p$) & T1 & T2\\ 
 \hline
 $x_{TS}$& 300 & 3000  \\ 
 $y_{TS}$ & 300 & 300\\
 \hline
\end{tabular}
\caption{Dimensions of the test section (TS); cf. \Cref{fig:mnras0_3}.}
\label{table:TS_dim}
\end{table}

\par
\begin{figure*}[h]
\centering
\includegraphics[width=\textwidth]{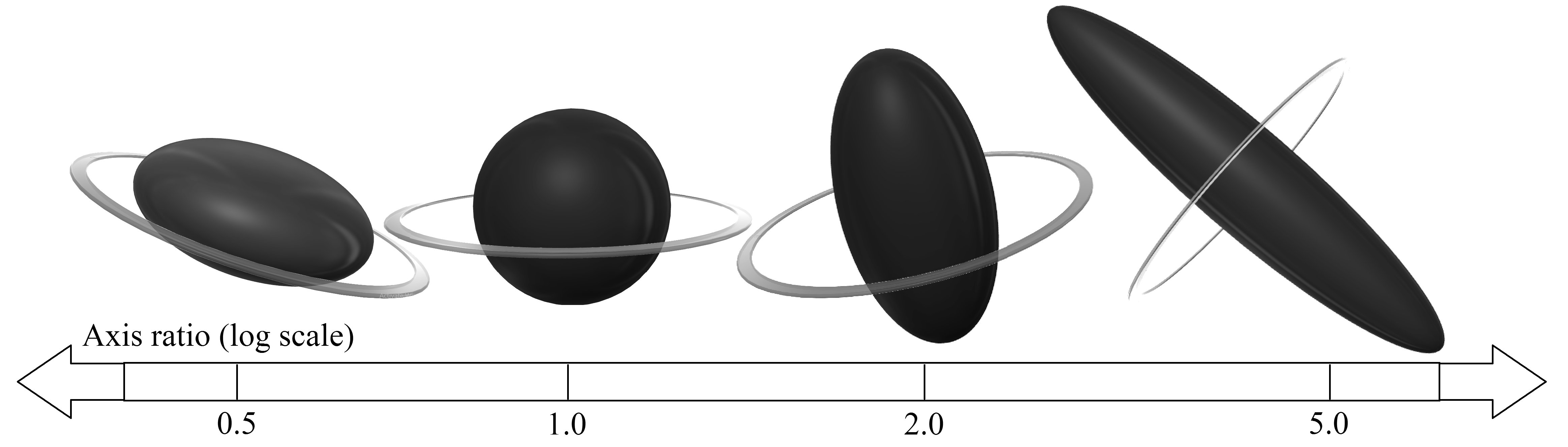}
\caption{Variation of central body shape with axis ratio $\theta=a_z/a_r$.}
\label{fig:mnras0_2}
\end{figure*}

 \Cref{table:sim_char} lists all important simulation parameters. All simulations reported in \Cref{sec:ell_res} are discussed in terms of the axisymmetric ellipsoidal central body's axes ratio
\begin{equation*}
\theta = a_z/a_r;
\end{equation*}
see \Cref{fig:ellipsoid_TS,fig:mnras0_2}. Mass and density of the central body are kept same across all simulations. Simulations discussed in  \Cref{sec:char_freq_var} are characterized by the characteristic frequencies $\kappa$, $\Omega$ and $\nu$; defined by (\ref{eq:eqm456}).

\par All our simulations have the same initial spatial distribution of particles with the center of mass of the ring particles lying at the TS' `guiding center' \textit{C}. The initial velocities of the particles are assumed to be Keplerian, found without considering the self-gravity. We follow previous simulation studies, e.g. \citet{wisdom1988a}, and assume that initial conditions do not affect the final equilibrium/steady-state. 

\begin{table}[h]
\centering
\begin{tabular}{c  c} 
 \hline
 Simulation parameter & Value / Range\\ 
 \hline
 Ellipsoidal central body's axis ratio ($\theta = a_z/a_r$) & $0.4-5.0$\\ 
 Mass of central body ($m_{cb}$) & $5.68 \times 10^{26}\;$kg\\
 Density of central body ($\rho_{cb}$) & $687\;$kg m$^{-3}$\\
 TS'distance from central body ($r_{gc}$) & $100,000\;$km\\ 
 Number of particles & $1432,\;14324^{\#}$ (identical) \\ 
 Initial optical depth ($\tau_0$) & $0.05,\;0.5^{\#}$\\
 Density of particle ($\rho_{p}$) & $900\;$kg m$^{-3}$ (solid-ice)\\
 Particle radius ($r_p$) & $1\;$m\\
 Coefficient of restitution ($e_n$) & $0.1$ (constant)\\
 Integration time-step ($dt$) & $2.5\;$s \\
 Total simulation time & $2.5\times10^7 \;$s, $2.5\times10^6 \;$s$^{\#}$\\ 
 \hline
\end{tabular}
\caption{Simulation parameters. The central body's mass and density match that of Saturn. $^{\#}:$ These parameters are employed only when investigating variation in the Toomre parameter.}
\label{table:sim_char}
\end{table}

\par Next, we discuss results from our simulations in terms of the local granular temperature, pressure, effective vertical thickness, effective radial width of the ring and impact frequency, which are defined in \Cref{table:prop_abb}. Definitions of these quantities and the manner in which they are computed is available in \ref{par_study}. All reported quantities are averaged over an extended period of time after the system has reached steady-state. 

We note that {\em all} plots of the main text are semi-log plots, with the abscissa carrying quantities on a log scale.

\begin{table}[h]
\centering
\begin{tabular}{ c  c  c  c } 
 \hline
 Properties & Symbols & Properties & Symbols \\ 
 \hline
 Granular temperature & $T_g$ & Velocity dispersion in $j^{th}$-direction  & $c_j$ \\
 Impact frequency & $\omega_c$ & Optical depth & $\tau$ \\ 
 Number density & $n_s$ & Effective vertical thickness & $t_v$ \\
 Effective radial width & $w_r$ & Effective spreading rate & $s_{w_r}$ \\
 Toomre parameter & $Q$ & Roche radius & $r_{Roche}$ \\
 Total pressure at $k=0$ plane & $p_{k}$ & Collisional and streaming pressure at $k=0$ plane & $p_{k,coll}$ and $p_{k,str}$ \\
 \hline
\end{tabular}
\caption{Various ring properties and their symbols. $j$ = \{$x,y,z$\}, $k$ = \{$x,z$\}.}
\label{table:prop_abb}
\end{table}

\section{Simulation results: Rings of axisymmetric ellipsoids} \label{sec:ell_res}
\par We now discuss results of T1 and T2 simulations for particles with (self-gravitating simulations, indicated by `SG') or without mutual-gravitation (non-gravitating simulations, represented by `NG').

\subsection{T1 simulations} \label{T1_sim_res}
\par
\begin{figure*} [h!]
\centering
\adjincludegraphics[width=\textwidth,trim={{0.075\width} {0.50\height} {0.075\width} {0.0\height}},clip]{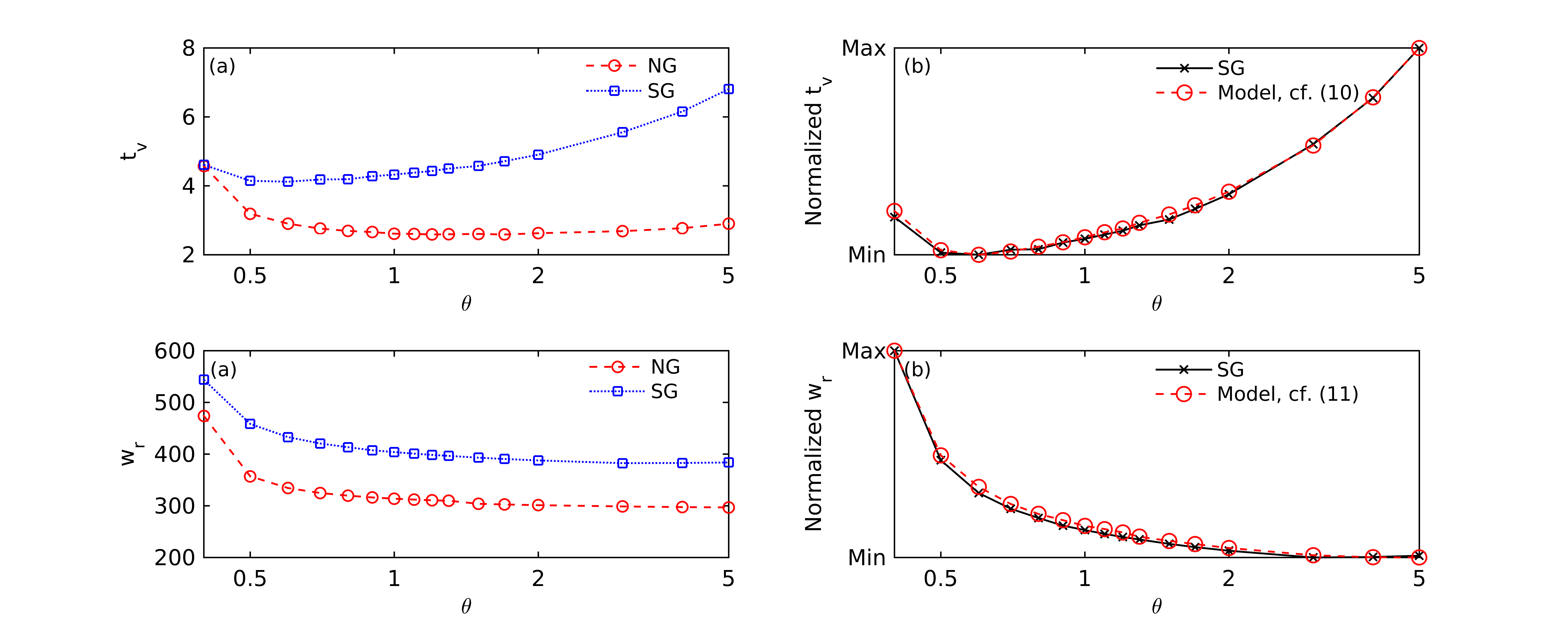}
\caption{T1 simulations. Variations in (a) effective vertical thickness $t_v$ (in particle radii) of the ring for non-gravitating (NG) and self-gravitating (SG) simulations with the axes ratio $\theta$, and (b) comparison of scaling predicted by simple model discussed in  \Cref{T1_sim_res} with simulation results. The vertical thickness is normalized by its maximum and minimum values. 
}
\label{fig:ellipsoid_par5}
\end{figure*}

 \Cref{fig:ellipsoid_par5,fig:ellipsoid_par4_1,fig:ellipsoid_par1} plot the variation of effective vertical thickness ($t_v$), granular temperature ($T_g$) and impact frequency ($\omega_c$), with the axes ratio $\theta=a_z/a_r$ of the central body. The Toomre parameter ($Q$) is considered in a subsequent section; cf. \Cref{fig:ellipsoid_par_t}. Simulations with non-gravitating and self-gravitating particles have been considered separately. We now discuss \Cref{fig:ellipsoid_par5,fig:ellipsoid_par4_1,fig:ellipsoid_par1} in greater detail.

\begin{enumerate}

\item \textit{Effective vertical thickness ($t_v$) and number density ($n_s$)}: We note that $t_v$  and $n_s$ are inversely proportional to each other for T1 simulations. We plot only the former in \Cref{fig:ellipsoid_par5}. We observe that as $\theta$ increases, $t_v$ initially decreases, reaches a minimum and then increases. The extrema occurs at a lower axes ratio for gravitating simulations. Interestingly, the ring's vertical thickness $t_v$ was found in self-gravitating simulations to be greater than that found when self-gravity was not active\footnote{\citet{salo1995a} also observed increased $t_v$ in SG simulations of a spherical central body. }, and this difference grew as the central body became more prolate. This is somewhat counter-intuitive, as we expect that in the presence of self-gravity particles are more attracted towards the equatorial plane, thereby reducing $t_v$.  We next explain this behavior through a simple model.
\par The fundamental idea is analogous to how equilibrium in stars is characterized in terms of a balance between thermal pressure and gravitational attraction \citep{shu1982a}. We consider vertical equilibrium at the equatorial plane ($z = 0$) of the ring. The vertical force per unit mass due to the central body's gravity, at any elevation $z$ from the equatorial plane, may be estimated from (\ref{eq:eqm3}) to be $-\nu^2 z$. Multiplying this by the mass $m$ of a particle and the number density $n_s$, and then integrating from $z = 0$ to $z = t_v$ yields the net force per unit area of the equatorial plane on the top half of the ring due to the central body's attraction: 
\begin{equation*}
	F_{cb} = -\frac{mn_s}{2}\nu^2 t_v^2, 
\end{equation*}
where the negative sign indicates that $F_{cb}$ points towards the equatorial plane. 

\par At the same time, the pressure exerted by the bottom half of the ring on its top half, due to particle transport across the equatorial plane  -- the streaming contribution -- may be estimated as follows.  The vertical momentum carried into the top half of the ring by particles per unit time and per unit area of the equatorial plane is 
\begin{equation*}
	\int_{v_z = 0}^{v_z = \infty} m v_z^2 f dv_z = mn_s c_z^2,
\end{equation*}
where $f$ represents a velocity distribution of the particles and $c_z$ is the vertical velocity dispersion. Similarly, because of symmetry, $mn_s c_z^2$ amount of vertical momentum flows per unit time across per unit area of the equatorial plane from the top half of the ring to the the bottom half. Thus, the total streaming pressure exerted by the bottom half of the ring onto the top half is 
\begin{equation*}
	p_{z, str} = 2 mn_s c_z^2.
\end{equation*}
For the equilibrium of the top half of the ring, $F_{cb}$ must balance $p_{z,str}$, which provides the scaling 
\begin{equation}
	t_v \propto \frac{c_z}{\nu}.
\end{equation}
\Cref{fig:ellipsoid_par5}(b) shows that the above relationship captures the observed behavior in simulations. 

At a fixed axes ratio, the inverse scaling of $t_v$ with the vertical frequency $\nu$ has been previously observed \citet{wisdom1988a}. However, we also see that $t_v$ is directly proportional to the velocity dispersion $c_z$. This helps explain the fact that $t_v$ is greater when self-gravity is active. In the presence of self-gravity, both $\nu$ and $c_z$ increase. The latter because self gravity will increase collisional velocities. It so happens that the effect of $c_z$ dominates.

\par An implicit assumption in the simple model above is that the primary contribution to the pressure is due to the streaming of particles across a plane. We numerically examine the validity of this assumption. \Cref{fig:ellipsoid_par4_1} shows the individual contributions to the total pressure $p_{z}$ at the $z=0$ plane from the streaming $p_{z,str}$ and collisional $p_{z,coll}$ components. It is clear that the streaming component dominates at most axes ratios $\theta$. This is expected, because the optical depth for the ring system is low. In self-gravitating simulations, the total pressure is modified by self-gravity which brings particles together. For prolate ($\theta>1$) bodies, the individual contributions become comparable in magnitude, although here too the total pressure nearly coincides with the streaming component, particularly at lower $\theta$. Finally, we note that the trend in the collisional component of the pressure is similar to what is observed for impact frequency in \Cref{fig:ellipsoid_par1}, that is discussed next.

\begin{figure} [h!]
\centering
\adjincludegraphics[width=\textwidth,trim={{0.075\width} {0.50\height} {0.075\width} 0},clip]{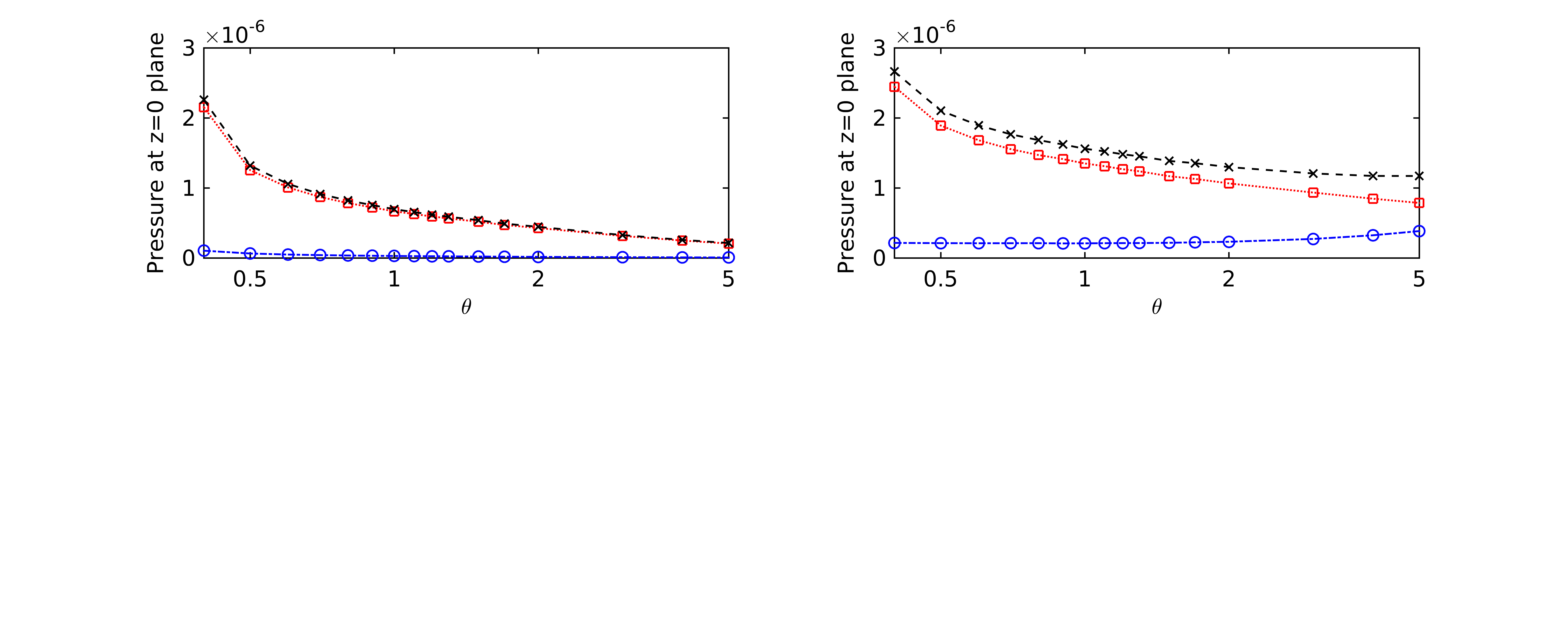} 
\caption{T1 simulations. Variations with the axes ratio $\theta$ in pressure at $z=0$ plane (in N/m$^2$) exerted by bottom half of ring on its top half for (a) non-gravitating and (b) self-gravitating simulations. Symbols correspond to the different components of the total pressure at $z=0$ plane: `{\large\color{blue}$\cdot$-{$\circ$}-$\cdot$}' for collisional component ($p_{z,coll}$), `{\color{red}$\cdot\cdot\square\cdot\cdot$}' for streaming component ($p_{z,str}$) and `-$\times$-' for the total pressure ($p_{z}$).}
\label{fig:ellipsoid_par4_1}
\end{figure}

\begin{figure} [h!]
\centering
\adjincludegraphics[width=\textwidth,trim={{0.075\width} {0.50\height} {0.075\width} {0.025\height}},clip]{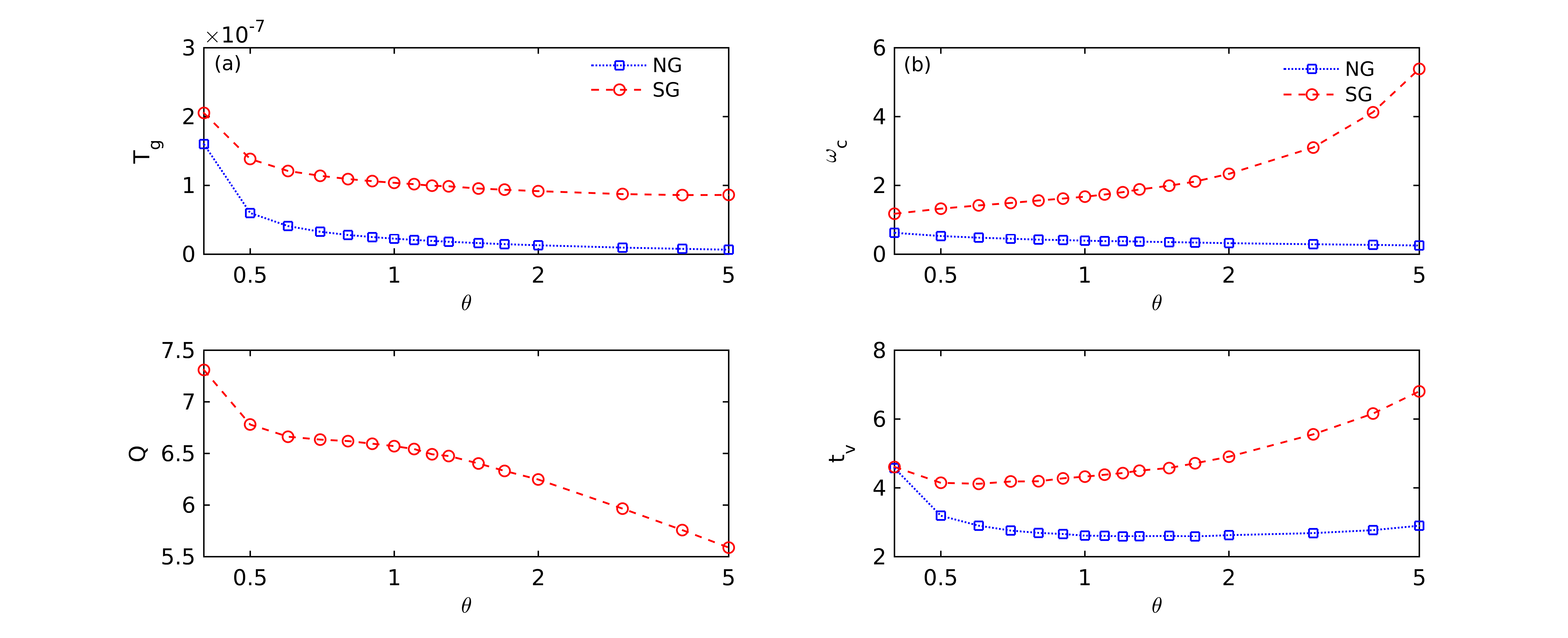}
\caption{T1 simulations. Variations in (a) granular temperature ($T_g$, {in m$^2$/s$^{2}$}) and (b) impact frequency ($\omega_c$, in s$^{-1}$) with the axes ratio $\theta$.}
\label{fig:ellipsoid_par1}
\end{figure}

\item \textit{Granular temperature ($T_g$)}: \Cref{fig:ellipsoid_par1}(a) shows that the local granular temperature -- a measure of the fluctuational kinetic energy -- decreases monotonically with increase in the axes ratio $\theta$, for both gravitating and non-gravitating ring. The ring is cooler when the central body becomes more prolate. The temperature $T_g$ is more sensitive to oblateness ($\theta < 1$), especially for $\theta<0.5$, and flattens out as the central body becomes prolate. 
This feature is common to both T1 and T2 simulations, with or without self-gravity.

\item \textit{Impact frequency ($\omega_c$)}: \Cref{fig:ellipsoid_par1}(b) shows the change in $\omega_c$ with axes ratio $\theta$ for gravitating and non-gravitating simulations. It is observed that including self-gravity increases $\omega_c$ for the same $\theta$. We discuss in \Cref{sec:behavior_char_freq} that, as $\theta$ increases, the influence of the central body's gravity decreases and, consequently, self-gravity gains more relative importance. Hence, as $\theta$ increases the difference between gravitating and non-gravitating rings also grows. 

\end{enumerate}

\subsection{T2 simulations}\label{T2_sim_res}
The primary aim of T2 simulations is to investigate the variation in the radial width ($w_r$) of the ring. In addition, we also report the qualitative behavior of the properties investigated in T1 simulations.

\begin{enumerate}
\item \textit{Effective radial width ($w_r$)}:  \Cref{fig:ellipsoid_par4}(a) shows that $w_r$ decreases as the axes ratio $\theta$ increases, with variations being much more noticeable at low $\theta$. This behavior is observed irrespective of the presence or absence of self-gravity. We also find that $w_r$, like $t_v$, is greater in self-gravitating simulations. It should be noted that the ring spreads radially continuously, although at the nominal steady state the spreading rate is nearly constant. The radial width is computed at a time when the vertical thickness has equilibrated. 
\par It is possible to construct a simple qualitative model to understand the behavior of $w_r$, as was possible for the vertical thickness. Applying a similar thermal pressure-gravity balance in the radial direction, we obtain  
\begin{equation}
{w_r} \propto c_x/\kappa,
\end{equation}
where $c_x$ is the radial velocity dispersion. \Cref{fig:ellipsoid_par4}(b) shows that this model fits the simulation results very well.

\par \Cref{fig:ellipsoid_par3} presents the plots for total pressure $p_x$ at $x=0$ plane and its collisional $p_{x,coll}$ and streaming $p_{x,str}$ components. As before, the streaming component of the pressure is of primary importance. In non-gravitating simulations, the contributions towards the total pressure from both streaming and collisional components decreases as the axes ratio $\theta$ increases. On the other hand, in self-gravitating systems, the magnitude of individual contributions are higher than their non-gravitating counterparts. At the same time, the collisional contribution to pressure appreciates, while that of the streaming component lowers as $\theta$ increases.

\begin{figure} [h!]
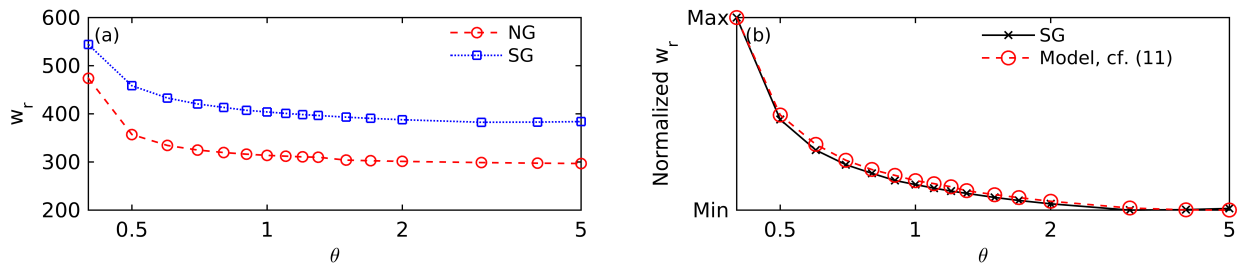

\centering
\adjincludegraphics[width=\textwidth,trim={{0.075\width} {0.025\height} {0.075\width} {0.50\height}},clip]{thickness_width_estimates3.png} 
\caption{T2 simulations. Variations in (a) effective radial width $w_r$ (in particle radii) of the ring for non-gravitating (NG) and self-gravitating (SG) simulations with the axes ratio $\theta$, and (b) comparison of scaling predicted by a simple model discussed in  \Cref{T2_sim_res} with simulation results. The radial width is normalized by its maximum and minimum values.}
\label{fig:ellipsoid_par4}
\end{figure}

\begin{figure} [h!]
\centering
\adjincludegraphics[width=\textwidth,trim={{0.075\width} {0.5\height} {0.075\width} 0},clip]{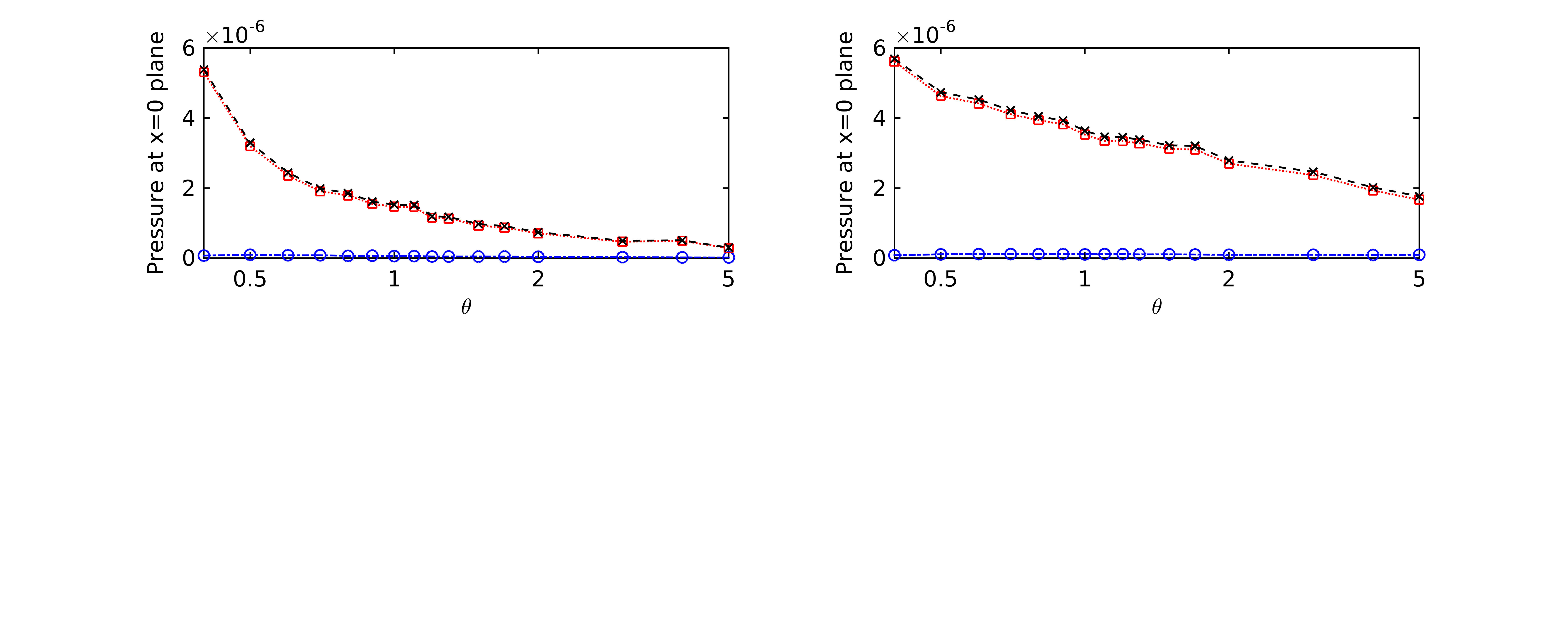}
\caption{T2 simulations. Variations with the axes ratio $\theta$ in pressure at $x=0$ plane (in N/m$^2$) exerted by the radially interior half of ring on its exterior half for (a) non-gravitating and (b) self-gravitating simulations. Symbols correspond to the different components of the total pressure at $x=0$ plane:  `{\large\color{blue}$\cdot$-{$\circ$}-$\cdot$}' for collisional component ($p_{x,coll}$), `{\color{red}$\cdot\cdot\square\cdot\cdot$}' for streaming component ($p_{x,str}$) and `-$\times$-' for the total pressure ($p_{x}$).}
\label{fig:ellipsoid_par3}
\end{figure}
\begin{figure}[h!]
\centering
\adjincludegraphics[width=\textwidth,trim={{0.075\width} {0.00\height} {0.075\width} {0.5\height}},clip]{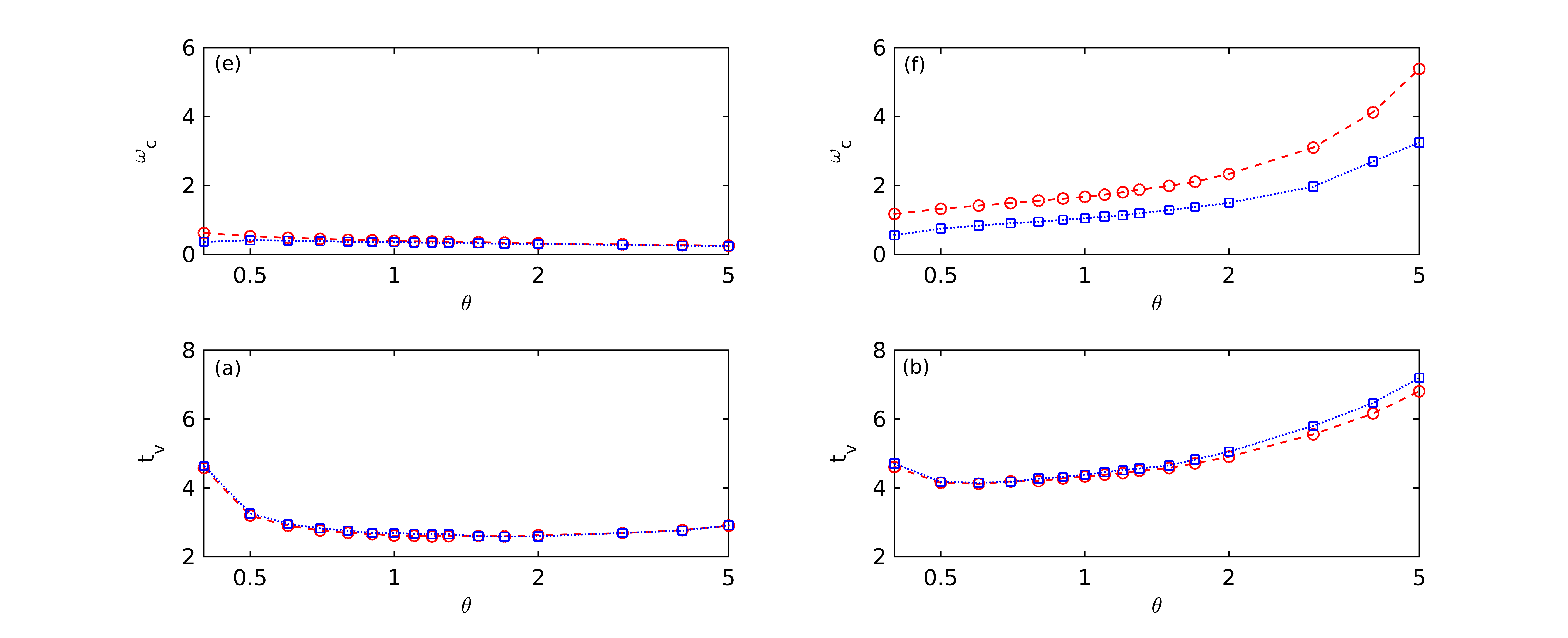}
\adjincludegraphics[width=\textwidth,trim={{0.075\width} {0.50\height} {0.075\width} {0.025\height}},clip]{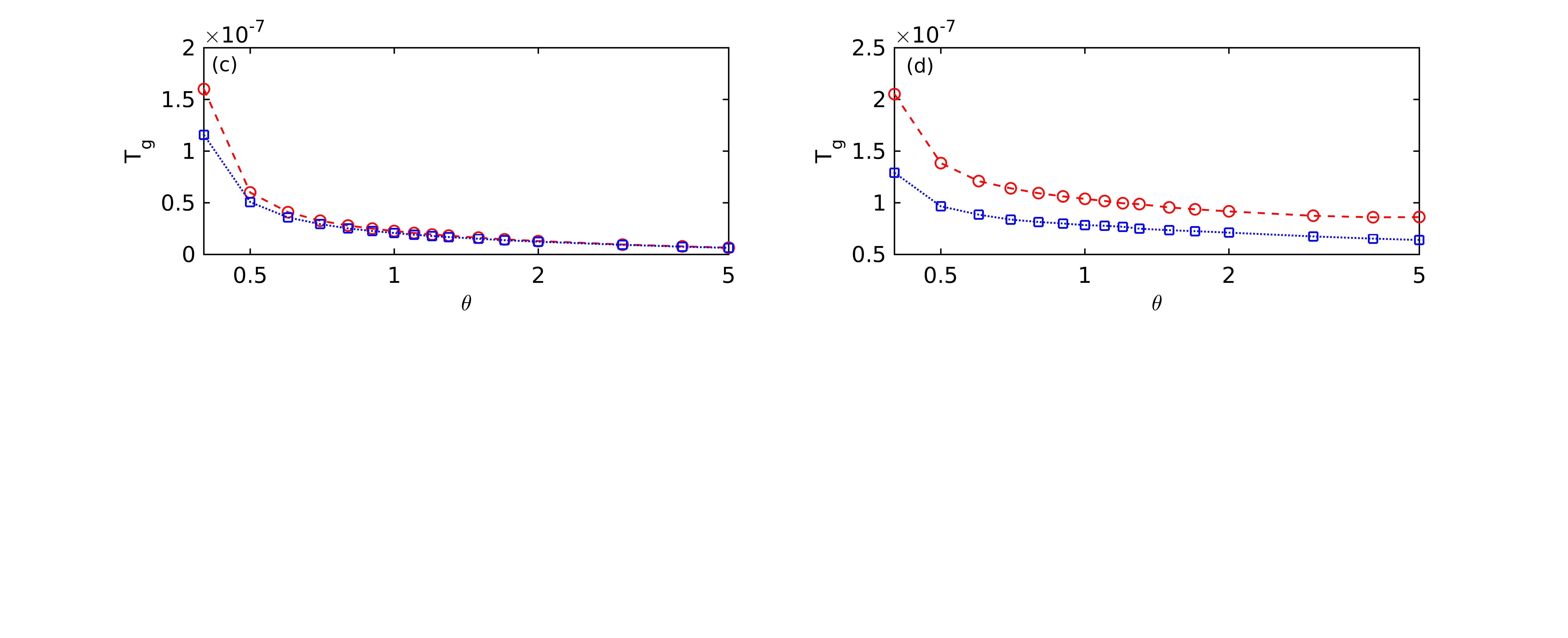}
\adjincludegraphics[width=\textwidth,trim={{0.075\width} {0.50\height} {0.075\width} {0.025\height}},clip]{t1vt2_1_new.png}
\caption{{Comparison of T1 ({\large \color{red}-$\circ$-}) and T2 ({\color{blue}$\cdot\cdot \square\cdot \cdot$}) simulations. Variations in (top row) effective vertical thickness ($t_v$, in particle radii), (middle row) granular temperature ($T_g$, in {m$^2$/s$^{2}$}) and (bottom row) impact frequency ($\omega_c$, in s$^{-1}$) with the axes ratio $\theta$. Left and right columns report results from non-gravitating and gravitating simulations, respectively.}}
\label{fig:T1vsT2}
\end{figure}
\item \textit{Other ring properties}: \Cref{fig:T1vsT2} compares predictions of T1 and T2 simulations for effective vertical thickness $t_v$, granular temperature $T_g$ and impact frequency $\omega_c$. Both non-gravitating (left panels) and self gravitating (right panels) cases are considered. 
\par We observe that in non-gravitating T1 and T2 simulations, $t_v$, $T_g$ and $\omega_c$ are nearly the same. Even when self-gravity is included, $t_v$ in T1 and T2 simulations remains about the same. However, $T_g$ and $\omega_c$ are significantly lower in self-gravitating T2 simulations. This maybe explained by noting that when self gravity is included in T2 simulations, particles spread more in the radial direction for all axes ratios. Indeed, \Cref{fig:ellipsoid_par4} shows that the nominal steady-state value of the effective radial width $w_r$ is always greater than $300$ particle radii, which equals the $x$-dimension of the TS employed in T1 simulations. Thus, particles effectively occupy a larger area in T2 simulations and have a lower number density $n_s$. This, in turn, reduces the number of impacts and the velocity dispersion in the $x$-direction.

\end{enumerate}

\subsection{Emergence of self-gravity wakes}
\begin{figure} [h]
\centering
\adjincludegraphics[width=\textwidth,trim={{0.075\width} {0.50\height} {0.075\width} 0},clip]{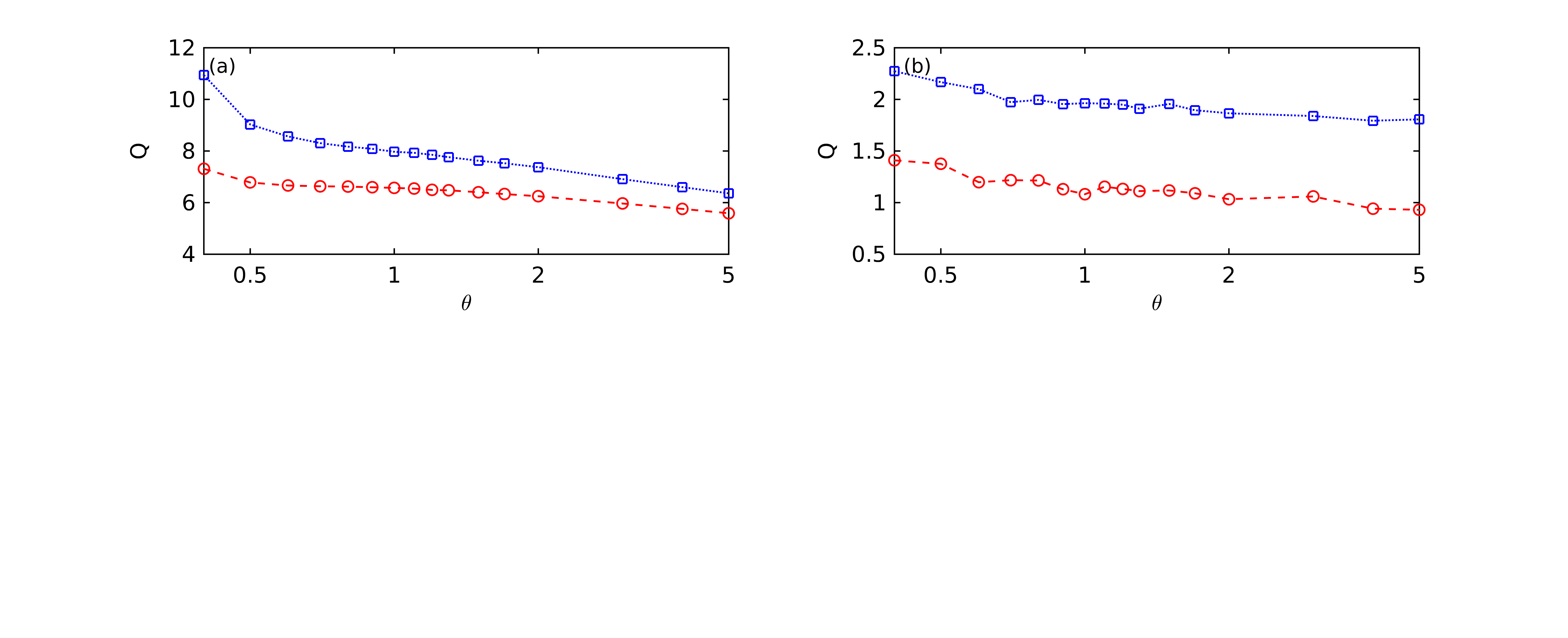}
\caption{Variation in Toomre parameter ($Q$) with the axes ratio $\theta$. Figure (a) corresponds to simulations with $\tau_0=0.05\;(1432 \textrm{ particles})$, while (b) $\tau_0=0.5\;(14328 \textrm{ particles});$ all other parameters are same as in \Cref{table:sim_char}. T1 ({\large \color{red}-$\circ$-}) and T2 ({\color{blue}$\cdot\cdot \square\cdot \cdot$}) simulations are shown.}
\label{fig:ellipsoid_par_t}
\end{figure}

\begin{figure}[h]
\centering
\adjincludegraphics[width=\textwidth,trim={{0.00\width} {0.00\height} {0.0\width} 0},clip]{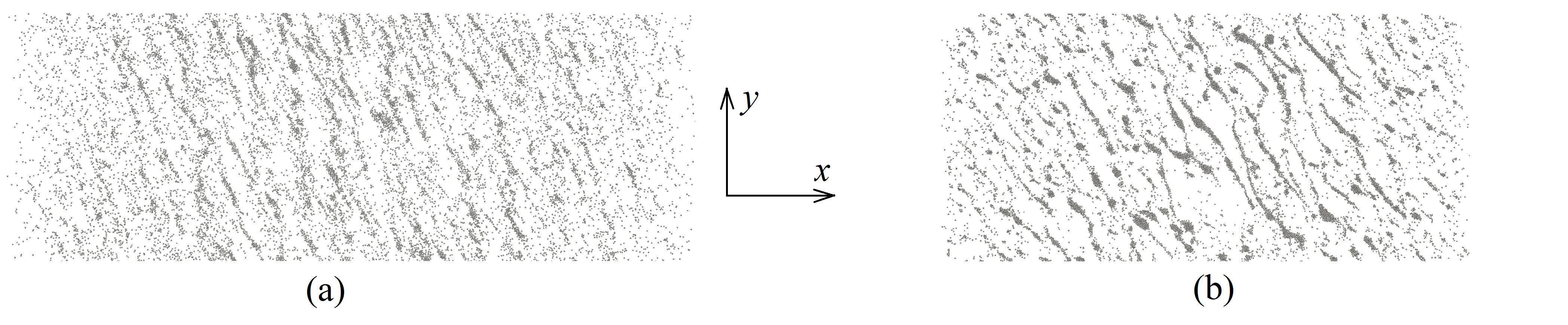}
\caption{Snapshots of T2 simulations for (a) $\theta$ = 0.5 (oblate), and (b) $\theta =  5.0$ (prolate). The initial optical depth was $\tau_0 = 0.5$. We find the a stronger presence of  self-gravity wakes in rings around more prolate bodies.}
\label{fig:mnras0_sg}
\end{figure}

Variations in the dynamical properties of rings due to changes in the central body's shape are also manifested by features/structures that could be found in rings about non-spherical bodies. As an example, we consider self-gravity wakes. The occurrence or strength of such instabilities may be quantified through the Toomre parameter $Q$, defined in \ref{par_study}. This parameter was computed for both T1 and T2 simulations, for two different initial optical depths $\tau_0$, and is shown in \Cref{fig:ellipsoid_par_t}. These plots show a decrease in $Q$ with increase in axes ratio $\theta$, indicating the growing susceptibility of the ring to axisymmetric disturbances, that may lead to self-gravity wakes. For the case of $\tau_0=0.05$, $Q$ is not low enough for these wakes to be manifested; see \Cref{fig:ellipsoid_par_t}(a). However, we find that wake structures are clearly observed  when $\tau_0=0.5$; see \Cref{fig:ellipsoid_par_t}(b). These wakes were stronger when the central body was prolate; see \Cref{fig:mnras0_sg}

\section{Rings of symmetric solids of revolution}

\begin{figure} [h]
\centering
\includegraphics[width=0.5\textwidth]{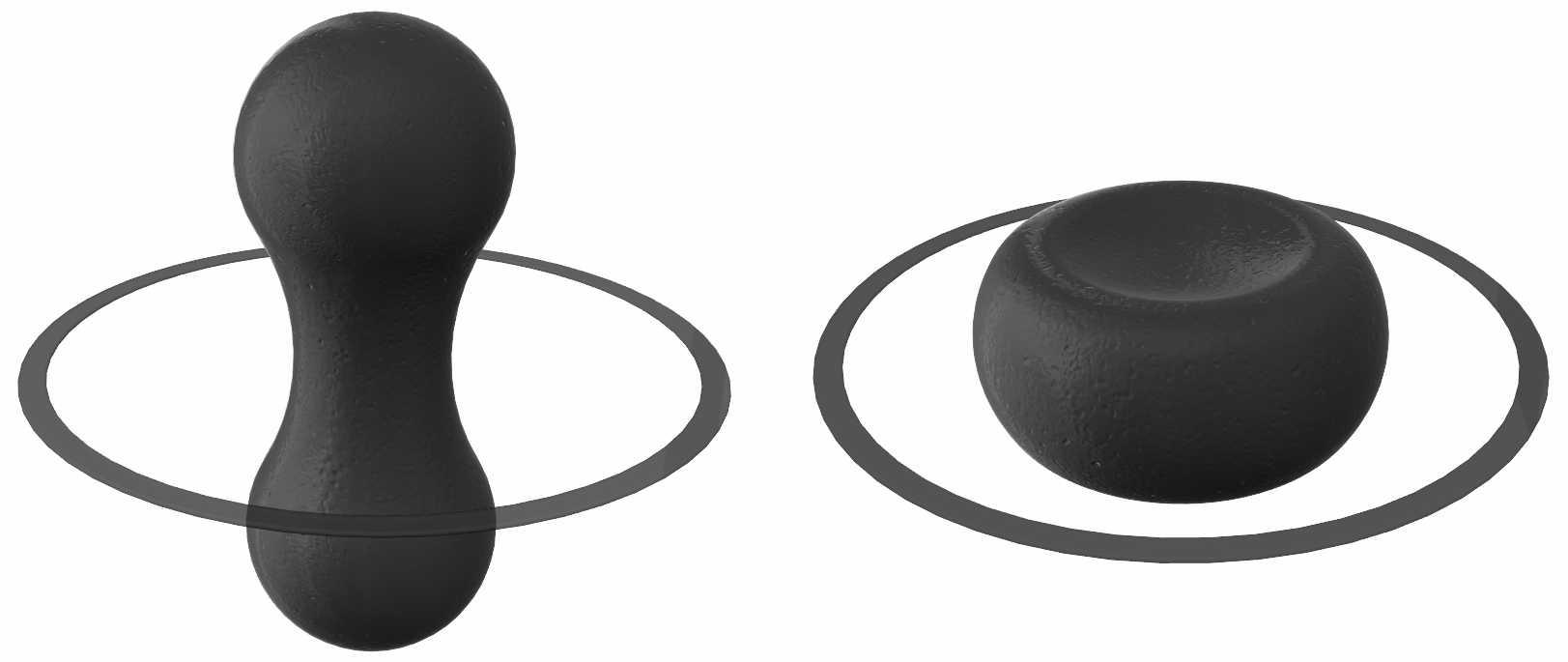} 
\caption{Two examples of symmetric solids of revolution with rings.}
\label{fig:axisymb}
\end{figure}

We are  investigating how the central body's shape affects the dynamics of its rings. In previous sections this was done by employing exact formulae for the external gravitational field of such aspherical bodies, restricting them to be symmetric ellipsoids. It is far more complex to repeat this exercise for non-ellipsoidal objects, as similar formulae for the gravity field are unavailable. In this section, we pursue an alternative approach to capture, in a qualitative manner, the effect that a non-spherical shape of the central body has on its rings. This approach rests upon the observation that a body's external field affects particle motion in the ring through only the characteristic frequencies $\kappa,\;\nu,\;\textrm{and}\;\Omega$, defined by (\ref{eq:eqm456}); cf. (\ref{eq:eqm123}). It is hoped that, once we understand how  characteristic frequencies relate to ring dynamics, it remains only to estimate these frequencies for the central body of interest; although that in itself can be complex. We will restrict ourselves to central bodies that are solids of revolution and have an equatorial plane of symmetry; two examples are shown in \Cref{fig:axisymb}. Our approach was inspired by past studies \citep{wisdom1988a} that mimicked  self-gravity in rings by simply augmenting the vertical oscillation frequency $\nu$ in non-gravitating local simulations,

%
%
%

\par First, in \Cref{sec:behavior_char_freq} we discuss how the characteristic frequencies vary as the axes ratio $\theta$ changes for symmetric ellipsoidal central bodies. We then discuss how the central body's shape, size and distance from the ring influence ring dynamics. In \Cref{sec:char_freq_var}, we investigate how the three characteristic frequencies individually influence ring dynamics. Finally, we discuss how we may qualitatively predict the dynamical properties of a ring about any solid revolution with an equatorial plane of symmetry.

\subsection{Characteristic frequencies} \label{sec:behavior_char_freq}
\par Equation (\ref{eq:eqm123}) shows that the non-sphericity of the central body influences particle dynamics through the three characteristic frequencies $\kappa,\;\nu,\;\textrm{and}\;\Omega$. These are, respectively, the epicyclic/radial frequency, vertical frequency and circular frequency. The parameter 
\begin{equation}
	\xi = ({\partial^2 \Phi}/{\partial r^2})_{r=r_{gc}, z=0} = (\kappa^2 - 3\Omega^2)_{r=r_{gc}, z=0},
\end{equation}
where $\Phi$ is the gravitational potential of the central body defined by (\ref{eq:ell_pot}), corresponds to the radial tidal force due to the central body in the equatorial plane.

\begin{figure*} [h]
\centering
\adjincludegraphics[width=\textwidth,trim={{0.075\width} {0.50\height} {0.075\width} 0},clip]{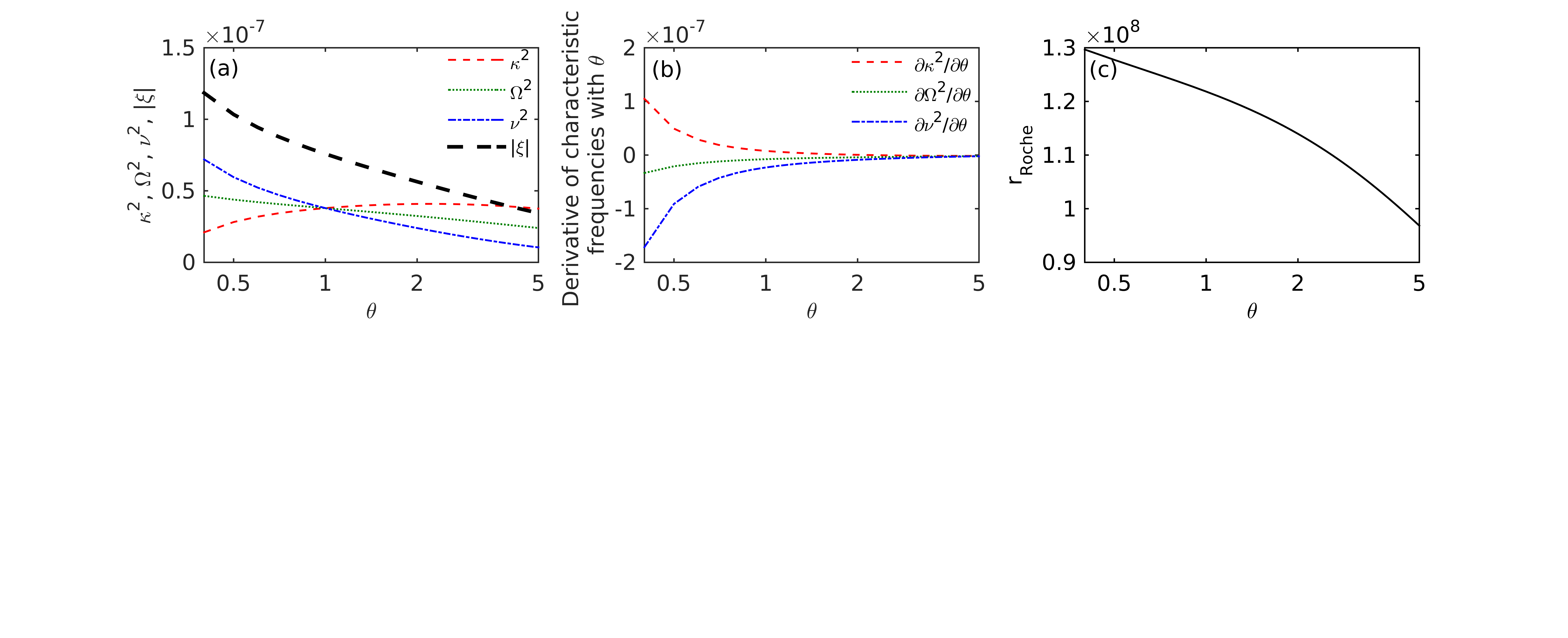}
\caption{Variations with axes ratio $\theta$ in (a) squares of characteristic frequency $\kappa^2$, $\Omega^2$, $\nu^2$ and $\xi$ (all in s$^{-2}$), (b) derivatives of characteristic frequencies, and (c) Roche radius ($r_{Roche}$, in m).}
\label{fig:frequencies_and_others}
\end{figure*}

\Cref{fig:frequencies_and_others} shows $\kappa,\;\nu,\;\Omega\;\textrm{and}\;\xi$ as function of the axes ratio $\theta$. Decreasing $|\xi|$ as $\theta$ increases indicates a lowering in the tidal forces in the radial direction at the center of the TS. Similarly, $\nu = |{\partial^2 \Phi}/{\partial z^2}|$, which is the gradient of the vertical gravitational force, and $\Omega$, which corresponds to the gravitational force of the central body in the radial direction, also reduce as the central body becomes less oblate (or more prolate). This means that the vertical oscillation frequency and the circular frequency are smaller for prolate bodies. On the other hand, the epicyclic frequency $\kappa$ grows initially, reaches a minimum and then decays. This non-monotonicity is because $\kappa$ is a sum of two terms; see (\ref{eq:eqm4}): both terms decrease in magnitude but have opposite signs. The occurrence and location of $\kappa$'s maximum depends upon system parameters, like $\theta$ and the distance of the TS from the central body. For a body like Saturn, the maximum occurs at $\theta$ $\sim 2$, while for a case like Chariklo we find no maximum and $\kappa$ steadily increases. 

\par The variation in the characteristic frequencies is less as the $\theta$ increases. This suggests a reduced significance of the central body and an increased relevance of inter-particle interactions when the system becomes more prolate. The growing importance of inter-particle forces, i.e., self-gravity and collisions, in comparison to the central body's gravity, in deciding the ring's steady-state properties may also be inferred by computing the Roche radius $r_{Roche}$. The Roche radius is the smallest distance from the central body at which two synchronously rotating particles persist in contact with each other due to mutual gravity; \ref{par_study} has a  quantitative definition and  formula (\ref{eq:rRoche}) helps compute $r_{Roche}$. \Cref{fig:frequencies_and_others}(c) shows that, as expected, $r_{Roche}$ decreases with increase in $\theta$.


\subsubsection{Effect of central body's size and radial distance of the ring}
\par The characteristic frequencies $\Omega$, $\kappa$ and $\nu$ may be simplified as

\begin{align*}
\Omega^2 = \beta \int^{\infty}_{\lambda(\theta, \zeta)} \frac{b(t,\theta)}{(1+t)}dt,\;\;
\kappa^2 = \beta \left[ 4 \int^{\infty}_{\lambda(\theta, \zeta)} \frac{b(t,\theta)}{(1+t)}dt - \frac{2}{\zeta(\zeta +\theta^{4/3} - \theta^{-2/3})^{0.5}} \right]\;\;\textrm{and}\;\;
\nu^2 = \beta \int^{\infty}_{\lambda(\theta, \zeta)} \frac{b(t,\theta)}{(\theta^2+t)}dt,
\end{align*}
where,
\begin{align*}
\zeta = \frac{r_{gc}^2}{(a_r^2 a_z)^{2/3}}, \;\; \lambda(\theta, \zeta) = \zeta \theta^{2/3} - 1  \;\; \textrm{and} 
\;\; b(t,\theta) = \frac{\theta}{(1+t)(\theta^2 + t)^{0.5}}.
\end{align*}
Thus, the three characteristic frequencies depend only upon two dimensionless quantities: axes ratio $\theta$ and the ratio $\zeta$ of the square of the distance of TS from the central body to the square of the effective size of the central body. Thus, for a given $\theta$, the dynamics of rings at a certain distance around a small sized central body will be similar to the dynamics found at a larger distance of a geometrically similar bigger body. This suggests that examining rings around a bigger body may provide insights into the rings of smaller similar shaped bodies.  Recent studies like \citet{braga2014a} follow this line of argument and expect the dynamical properties of Chariklo's rings to be akin to those observed in specific regions of other planetary-ring systems. This is because \citet{braga2014a} estimate the mean orbital frequency to be  $\Omega \sim 10^{-4}\;s^{-1}$, which is close to the orbital frequency of rings of Uranus and of the outer part of the rings of Saturn. However, this speculation is based on the assumption of a spherical Chariklo, which is {\em not} the case. A non-spherical shape affects the ring dynamics, as observed in \Cref{sec:ell_res} for axisymmetric ellipsoidal central bodies.

\par
\begin{figure} [h!]
\centering
\adjincludegraphics[width=\textwidth,trim={{0.075\width} {0.49\height} {0.070\width} {0.0\width}},clip]{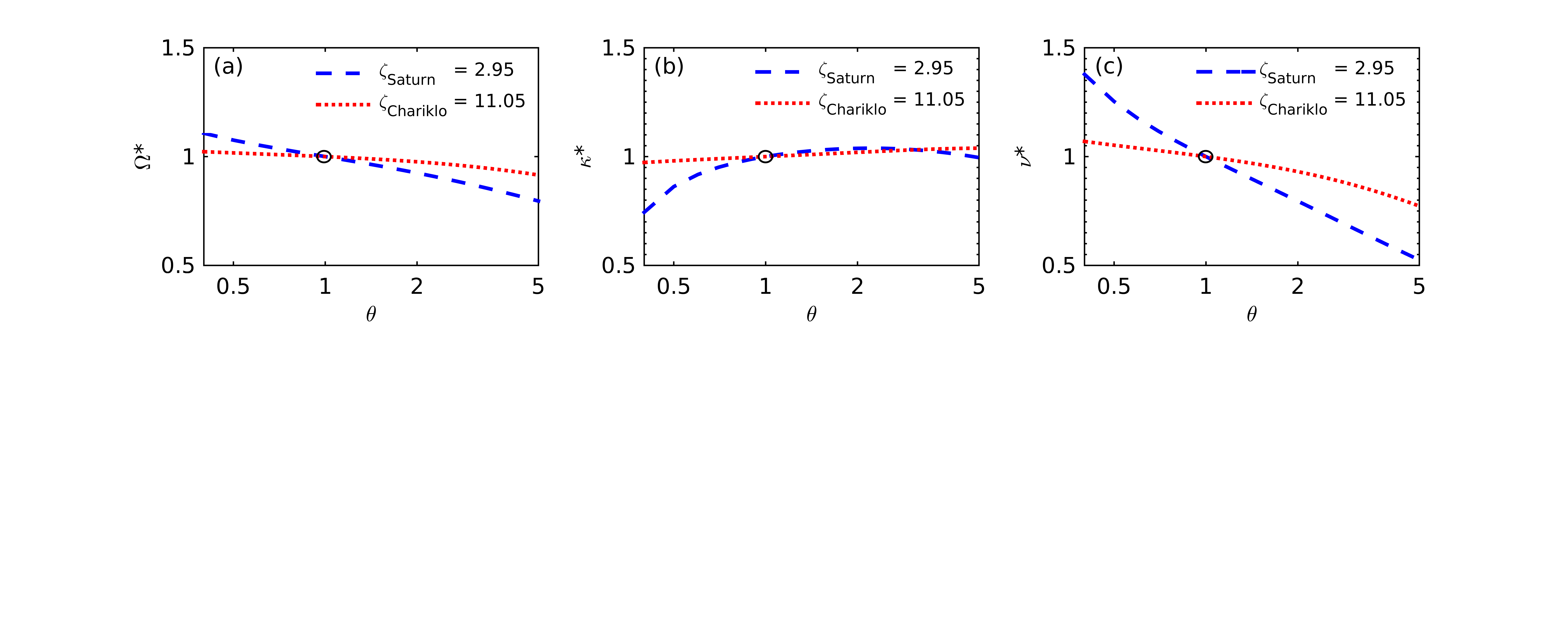}
\caption{Variations in ${\kappa^*},\;{\Omega^*}\; \textrm{and} \;{\nu^*}$ with axes ratio $\theta$ for two choices of $\zeta$.} 
\label{fig:zeta_senst}
\end{figure}

\begin{figure} [h!]
\centering
\adjincludegraphics[width=\textwidth,trim={{0.075\width} {0.35\height} {0.075\width} 0},clip]{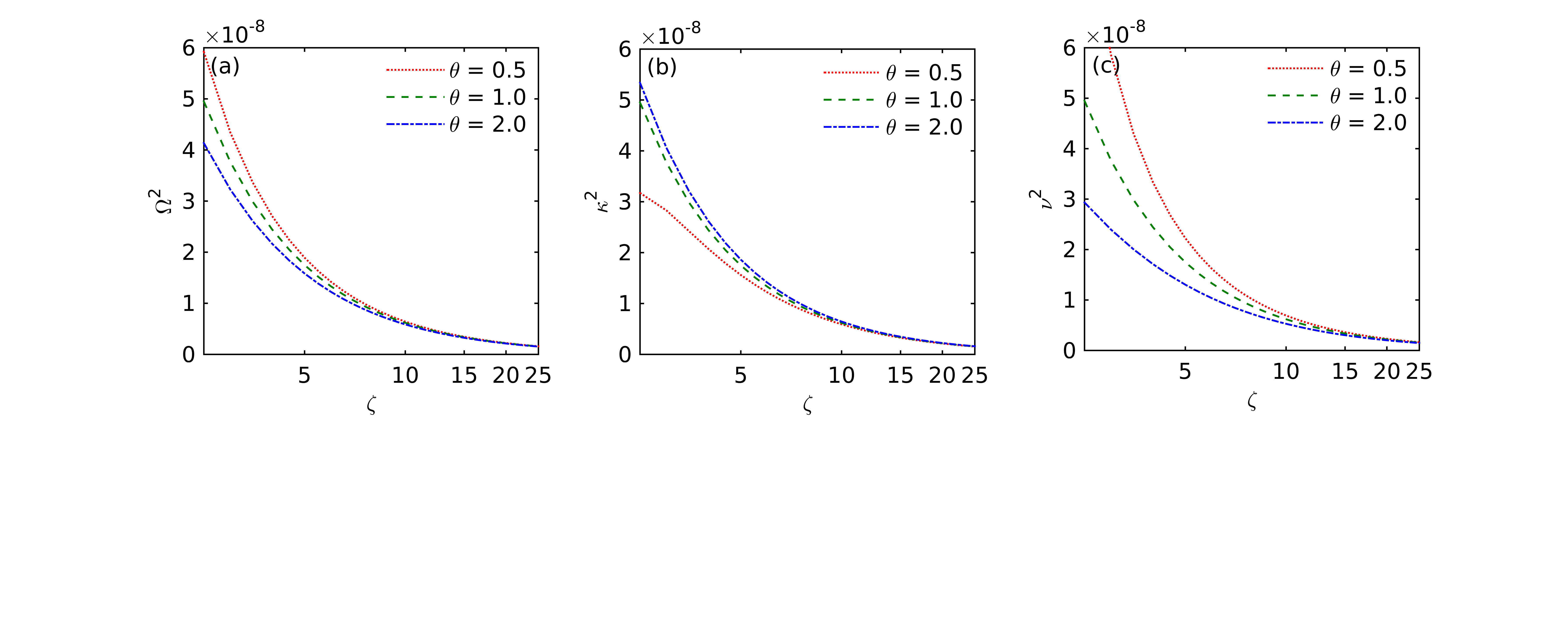} 
\caption{Variations in $\Omega^2$, $\kappa^2$ and $\nu^2$ with $\zeta$ at three axes ratios $\theta$.}
\label{fig:zeta_var}
\end{figure}

 For differently sized central body-ring systems, the variation in characteristic frequencies is shown in \Cref{fig:zeta_senst}, where the frequencies have been scaled as follows:

\begin{align}
\Omega^* = \frac{\Omega}{\Omega_0}, \; \; \kappa^*  = \frac{\kappa}{\kappa_0}   \; \; \textrm{and} \;\; \nu^*  = \frac{\nu}{\nu_0},
\label{eq:scaledfreq}
\end{align}
where the subscript `0' denotes frequencies obtained for a spherical central body with $\theta=1$. For a spherical Saturn, $\Omega_0=\kappa_0=\nu_0\sim1.94\;X\;10^{-4}\;$s$^{-1}$. In \Cref{fig:zeta_senst} two choices of $\zeta$ are explored. Saturn has an effective radius of $58,232$ km, so that a ring at $r_{gc} = 100,000$ km corresponds to $\zeta_{\textrm{Saturn}} = 2.95 $, whereas Chariklo, with equatorial radius taken to be $144.9\;$km, and oblatenes ($1 - a_z/a_r$) as $0.213$ \citep{braga2014a}, and a ring at $400$ km yields $\zeta_{\textrm{Chariklo}} =11.05$.  It is clear from \Cref{fig:zeta_senst} that the smaller body is less sensitive to local variations in the axes ratio $\theta$, which has implications for observations that seek to describe Chariklo's precise shape.  However, irrespective of $\zeta$'s value, there is little or no change in the qualitative behavior of the characteristic frequencies with $\theta$; cf. \Cref{fig:zeta_var}.

\Cref{fig:zeta_var} shows the three characteristic frequencies as a function of $\zeta$ for three values of $\theta$. All frequencies lower with $\zeta$. The effect of $\theta$ decreases with increase in $\zeta$. This is expected as, for a fixed shape of the central body, larger the $\zeta$, the further away is the ring relative to the central body's size and, consequently, lower is the effect of the central body's shape on its gravity field.

\subsection{Effect of characteristic frequencies} \label{sec:char_freq_var}
%
%

\par In this section, we first observe the influence upon the ring's properties of the characteristic frequencies $\Omega, \kappa$ and $\nu$. Then, employing the symmetric ellipsoid as an example, we demonstrate how combining this information with the knowledge of how characteristic frequencies depend on the axes ratio (\Cref{sec:behavior_char_freq}) will allow us to qualitatively reconstruct the trends observed in \Cref{sec:ell_res}. This suggests a way in which we may predict the qualitative features of rings of central bodies which are a solid of revolution with their symmetry axis normal to the equatorial plane, e.g., \Cref{fig:axisymb}, provided the higher-order corrections to the central body's gravitational potential remain small.

\begin{figure} [!h]
\centering
\adjincludegraphics[width=\textwidth,trim={{0.075\width} {0.50\height} {0.075\width} 0},clip]{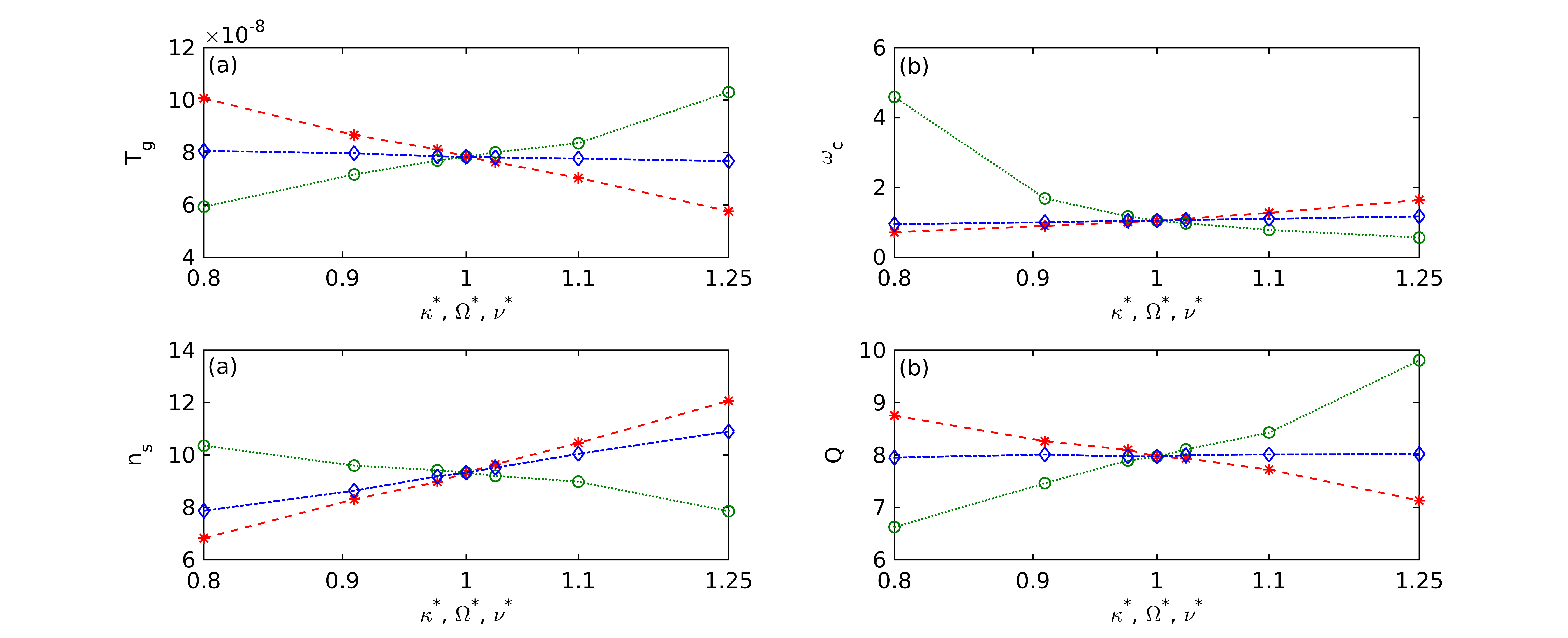} 
\adjincludegraphics[width=\textwidth,trim={{0.075\width} {0.5\height} {0.075\width} 0},clip]{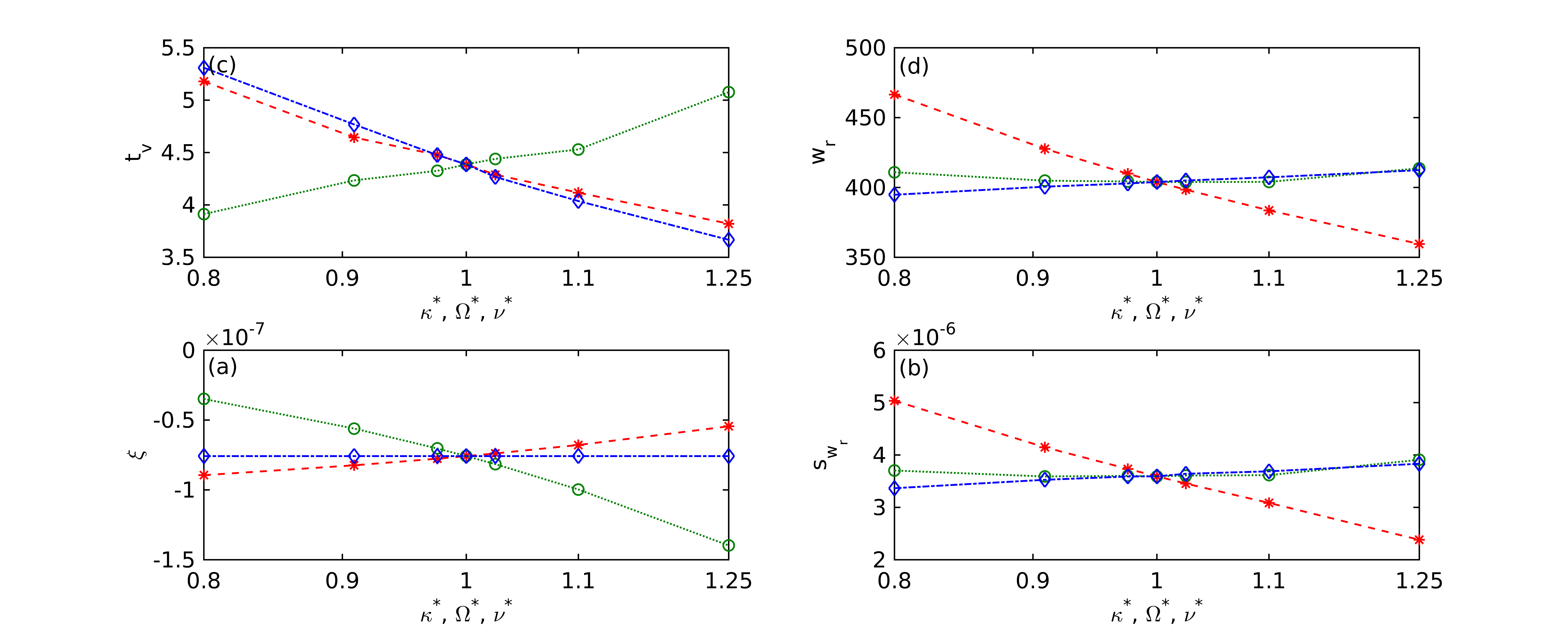} 
\caption{Self-gravitating T2 simulations: Variations with characteristic frequencies in (a) granular temperature ($T_g$, in {m$^2$/s$^{2}$}), (b) impact frequency ($\omega_c$, in s$^{-1}$), (c) effective vertical thickness ($t_v$, in particle radii), (d) effective radial width ($w_r$, in particle radii). Symbols correspond to the different characteristic frequencies: `{\color{red}- -$\ast$- -}' for $\kappa^*$, `{\color{green}$\cdot\cdot\circ\cdot\cdot$}' for $\Omega^*$ and `{\color{blue}$\cdot$ -$\diamond$- $\cdot$}' for $\nu^*$.}
\label{fig:T2_char_freq_par1}
\end{figure}

\Cref{fig:T2_char_freq_par1,fig:T2_char_freq_par2,fig:T2_char_freq_par4_1,fig:T2_char_freq_par4_2} plot various ring properties against scaled characteristic frequencies $\Omega^*, \kappa^*$ and $\nu^*$; cf. (\ref{eq:scaledfreq}). We have performed T1 and T2 simulations for only self-gravitating rings. The observed trends of T1 and T2 simulations are similar, and, we present and discuss T2 simulations alone. All aspects of the simulations follow the paradigm of \Cref{sec:sim_spec}. Note that the central body's gravitational potential need not be defined in these simulations, as we work directly with (\ref{eq:eqm123}) after selecting $\Omega, \kappa$ and $\nu$, which are treated as {\em parameters}.

\par We now discuss in some detail the trends observed in \Cref{fig:T2_char_freq_par1}. Furthermore, employing these plots with \Cref{fig:zeta_senst}, we reconstruct trends observed for T2 simulations of an symmetric ellipsoidal central body. 

\begin{enumerate}
\item Granular temperature ($T_g$): \Cref{fig:T2_char_freq_par1}(a) shows that the $T_g$ decreases with rising $\kappa^*$ and $\nu^*$, and increases with $\Omega^*$. However, $T_g$ is much less sensitive to changes in $\nu^*$. At the same time, \Cref{fig:zeta_senst} reported an appreciation in $\kappa^*$ and decay of $\Omega^*$ with growing $\theta$. Together, both trends indicate a net reduction in $T_g$ as the central body becomes more prolate (less oblate), as was observed in \Cref{fig:T1vsT2}(d). 

\item Impact frequency ($\omega_c$): \Cref{fig:T2_char_freq_par1}(b) shows that for $\Omega^*<1.0$, $\omega_c$ rises sharply with decreasing $\Omega^*$. This is primarily due to reduction in the radial tidal force $\xi$, which augments self-gravity driven collisions. Because $\Omega^*$ reduces with increasing $\theta$ in \Cref{fig:frequencies_and_others}(a), the sharp decay of $\omega_c$ with $\Omega^*$ dictates the initial positive change in $\omega_c$ with $\theta$  in \Cref{fig:T1vsT2}(f).
\par Similar arguments explain variations in $\omega_c$ due to radial oscillation frequency $\kappa^*$. In absence of self-gravity, the characteristic frequencies are the only factors dictating the response of $\omega_c$, observed in \Cref{fig:T1vsT2}(e). At higher frequencies ($>1.0$), the relative changes in $\omega_c$ with $\kappa^*$  and $\Omega^*$ are of comparable magnitude. It maintains a monotonic behavior, and increases with $\kappa^*$, but decreases with rising $\Omega^*$. 
\par Variations in $\omega_c$ with $\nu^*$ are relatively small, although a slight monotonic increase with $\nu^*$ is observed.

\item Effective vertical thickness ($t_v$): \Cref{fig:T2_char_freq_par1}(c) shows that $t_v$ decreases with $\kappa^*$ and $\nu^*$, and increases with $\Omega^*$. From \Cref{fig:zeta_senst} it is seen that for a Saturn-like object $\kappa^*$ grows with $\Omega^*$ and $\nu^*$ decay with $\theta$. Lowering $\Omega^*$ and augmenting $\kappa^*$ imply a reduction in $t_v$, whereas a decreasing $\nu^*$ implies an increasing $t_v$. Given their opposing trends, the final behavior of $t_v$ depends upon the relative sensitivity of $t_v$ on these frequencies.
\par At smaller values of $\theta$, variations in $\kappa^*$ and $\Omega^*$ are comparable in magnitude, but opposite in signs. Thus, together they overcome the opposing effect of $\nu^*$ and cause the initial decrease of $t_v$ with $\theta$, as seen in \Cref{fig:T1vsT2}(b). But, as $\theta$ increases further, the slope of $\kappa^*$ lowers sharply, with $\kappa^*$ beginning to reduce for $\theta>2$; see \Cref{fig:frequencies_and_others}(b). Moreover, because the relative change in $\nu^*$ is greater than that in $\Omega^*$, upon further increase in $\theta$, both these factor contribute to the subsequent upturn and growth of $t_v$, as observed in \Cref{fig:T1vsT2}(b). 

\item Effective radial width ($w_r$): \Cref{fig:T2_char_freq_par1}(d) shows that $w_r$ decreases with increase in $\kappa^*$, but grows with $\nu^*$. Unlike $t_v$, which depends comparably on all characteristic frequencies, $w_r$ depends significantly only on $\kappa^*$.  \Cref{fig:frequencies_and_others}(b) shows that $\kappa^*$ increases with $\theta$, so that we can conclude that $w_r$ reduces with the axes ratio, as was also  observed in \Cref{fig:ellipsoid_par4}(a).

\item Other properties: Observations from \Cref{fig:T2_char_freq_par2,fig:T2_char_freq_par4_1,fig:T2_char_freq_par4_2} are summarized in \ref{trends_char_freq}.
\end{enumerate}

\section{Conclusions}

\par In this work, we have investigated the effect of the shape of the central body on the dynamics of its rings. The central body was restricted to be an axisymmetric ellipsoid. An ellipsoidal shape was selected because closed form formulae are available for its external gravity field. Axisymmetry was assumed to allow us to perform local simulations with periodic boundary conditions. For simulations we developed a new discrete element (DE) code. Our code accounts for dissipative collisions between identical particles via discrete element method and also for gravitational forces between them.  We probed the effect of the central ellipsoid's axes ratio $\theta$ on various ring properties, e.g., granular temperature, effective vertical thickness, impact frequency, effective radial width, Toomre parameter, etc. All properties were affected by changes in $\theta$, especially when $\theta < 1$. System with or without self-gravity were simulated, and it was found that the differences between these two simulations depended on $\theta$. Simple models were developed to understand some of trends that we report. 

\par We then extended the scope of our investigation to central bodies that are solids of revolution with an equatorial plane of symmetry. This was done by understanding the effect on the ring's dynamical properties because of the characteristic frequencies governing epicyclic/radial, vertical and circular motion. The shape of the central body affects particle motion only through these characteristic frequencies. We also investigated how, apart from central body's shape, its size and the radial location of the ring affect the ring dynamics. Finally, we demonstrated how this may facilitate a qualitative understanding of the dynamics of rings of more general shapes. 

\par We show that a non-spherical gravitational potential affects the ring's dynamical properties and thus, because of Chariklo's non-spherical shape, it will be incorrect to expect -- as \citet{braga2014a} do -- that Chariklo's rings display behavior similar to regions of rings of the nearly spherical planets.


\par Our work indicates several new avenues of exploration. One, the development of a local simulation with appropriate boundary conditions to simulate rings of (non-axisymmetric) triaxial ellipsoidal central bodies. Next, it will be of interest to investigate the possibility of employing accurate ring simulations to estimate internal and geometric features of the central body from the observed ring properties. \citet{hedman2013a} have made some progress in this direction for understanding Saturn's interior. At the same time, it will be useful to extend kinetic theory models  of rings of spherical bodies \citep{jenkins2012a} to irregular shaped bodies. Further, it is required to couple our DE code with efficient computations of gravity fields of irregular bodies in order to simulate rings of  non-ellipsoidal central bodies. Finally, in the context of central bodies of general shape, it is not clear when rings may form and remain stable. This latter question gains in importance considering  that rings of small bodies may be a more common phenomenon than previously expected.

\section*{Acknowledgements}
A.G. is grateful to Prof. Heikki Salo, University of Oulu for guidance in developing his understanding of numerical simulations of planetary rings.
S.N. would like to acknowledge the Science and Engineering Research Board (SERB), Government of India for the grant (YSS/2014/000526).
I.S. would like to acknowledge the Planetary Science and Exploration (PLANEX) program from the Physical Research Laboratory, India for the grant (PRL/ADM-AC/PB/2013-14).
We would also like to acknowledge the IIT Kanpur High Performance Computing (HPC) facility.%

\section*{References}
  \bibliographystyle{elsarticle-harv} 
  \bibliography{asteroidring_icarus}




%

 \appendix
 
\section{Discrete element simulation: Details} \label{sim_detail}

\noindent In this appendix we provide details of our discrete element (DE) code in addition to those discussed in \Cref{sim_method}. We build upon the DE code developed by \citet{bhateja2014a} who, in turn, followed \citet{cundall1979a}. In DE simulations, particles are modeled as rigid bodies that interact with each other through collisions, which are, in turn, resolved after invoking an appropriate collisional model. Newton-Euler equations of motion are integrated to follow particle trajectories. Multiple, simultaneous collisions are permitted, which is an advantage over typical $N$-body codes and permits better modeling of agglomeration processes.  

\par A fourth-order Runge-Kutta (RK4) scheme is utilized for integrating equations of motion.  Particle configuration is updated efficiently though a `cell-list' technique  \citep{allen1987a}. This involves discretizing the entire test section (TS) into blocks, and tracking the movement of particles in and across these blocks. Gravity between ring particles is accounted through a similar `cell-list' structure, albeit with a different block size. Gravitational interactions are only considered between particles within a certain cut-off distance of each other. The size of a particle's  `sphere of influence' is proportional to its Hill's sphere. For simulations with identical particles this cut-off distance is set to fifty times the particle radius.

\par
\begin{figure}[h]
\centering
\includegraphics[width=0.5\columnwidth]{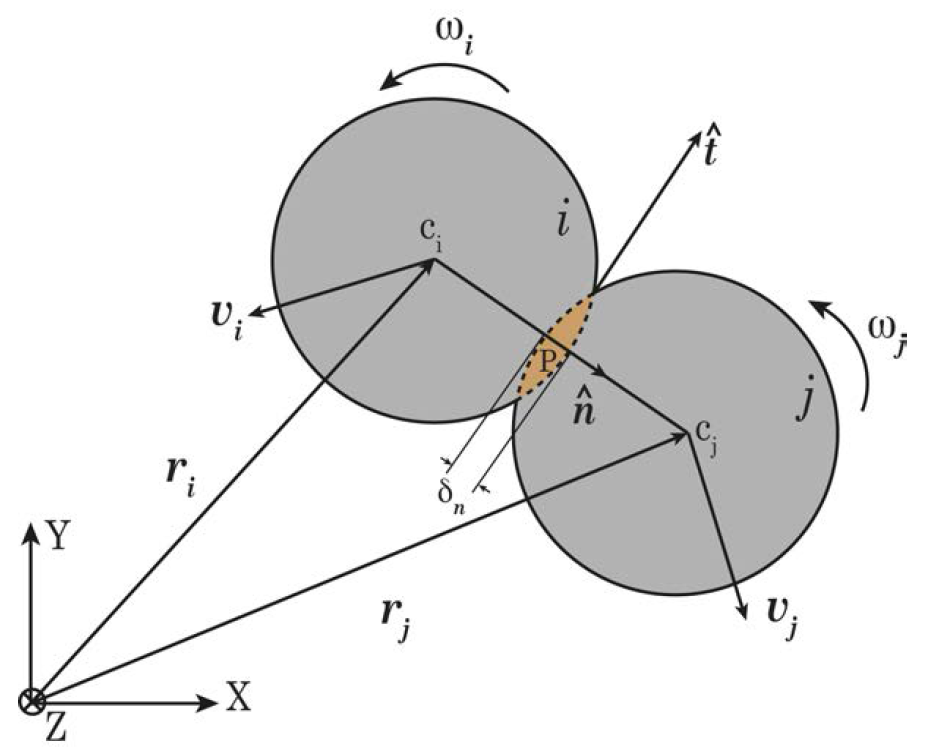} 
\caption{A schematic of the collision model between two particles in a two-dimensional representation. The normal $\hat{\textbf n}$ and tangential $\hat{\textbf t}$ directions are shown. Two particles interact if $|\textbf{r}_{ij}|$,  where $\textbf{r}_{ij} = \textbf{r}_j -\textbf{r}_i$, is less than $(R_i + R_j)$. The figure follows \citet{bhateja2014a}.}
\label{fig:mnras1}
\end{figure}

 The contact between two particles requires the specification of normal and tangential interactions. The normal direction is along the line joining centers of the two interacting particles, and  tangential directions are perpendicular to this. \Cref{fig:mnras1} shows the interaction of two particles `$i$', and `$j$'. \citet{cundall1979a} model the contact of two particles by first allowing particles to `overlap'; see \Cref{fig:mnras1}. It is assumed that these `overlaps' are small in comparison to the particle radii.  The magnitude of this overlap is then related directly to the contact force between the two particles through a combination of springs, dashpots and sliders. The spring stiffness ($k$), the damping coefficient ($c$) and coefficient of friction ($\mu$) capture a particle's microscopic properties of elasticity, collisional dissipation and surface friction.  Here we utilize a linear spring-dashpot model \citep{zhang1996a}. The spring is linear and Hookean, although a non-linear spring based on the Hertzian contact theory may also be implemented \citep{tsuji1992a}. The dashpot too is linear, while the friction slider follows Coulomb's law.  \citet{bhateja2016a} discusses further details of this approach and its implementation.

\par In our simulations we restrict ourselves to normal collisional interactions of smooth, non-spinning particles. The normal damping constant $c_n$ is determined from the normal coefficient of restitution $\epsilon_n$ that, in turn, may be estimated experimentally \citep{hatzes1988a,bridges1984a}. Studies show that $\epsilon_n$ is sensitive to factors like collisional velocities, surface properties, etc. For simplicity we employ a constant $\epsilon_n$. 

\par We  employ a modified version of a subroutine from the freely available QUADPACK library \citep{piessens1983a} for preforming semi-infinite integrations. This is a globally adaptive automatic quadrature algorithm, with the local quadrature module being a Gauss-Kronrod quadrature resting upon a 7-point Gauss and a 15-point Kronrod rule.

\par To make our code efficient we have utilized suggestions from earlier works on numerical simulation of planetary rings \citep{salo1995a, karjalainen2004a}. Thus, we update gravity calculations in an interval of $50  \delta t$, where $\delta t$ is the integration time-step. The spring constant is selected following \citet{salo1995a}. We employ `puffy' particles that have a relatively low spring constant. Thus, the time of collision is relatively larger than expected. This ensures that simulations spanning numerous orbits can be performed feasibly. Finally, we have included  second-order gravity corrections suggested by \citet{karjalainen2004a}.

\section{Code Validation} \label{code_val}
\par To validate our code we have made two-fold efforts. First, we ensured that all fundamental elements of our simulation provided valid results. This included the implementation of collisions, self-gravity between particles, central body force, $N$-body integrator and periodic boundary conditions. Many aspects, such as the collisional module, have been verified previously by \citet{bhateja2014a}.

\begin{figure}[!h]
\centering
\adjincludegraphics[width=\textwidth,trim={{0.075\width} {0.00\height} {0.075\width} 0},clip]{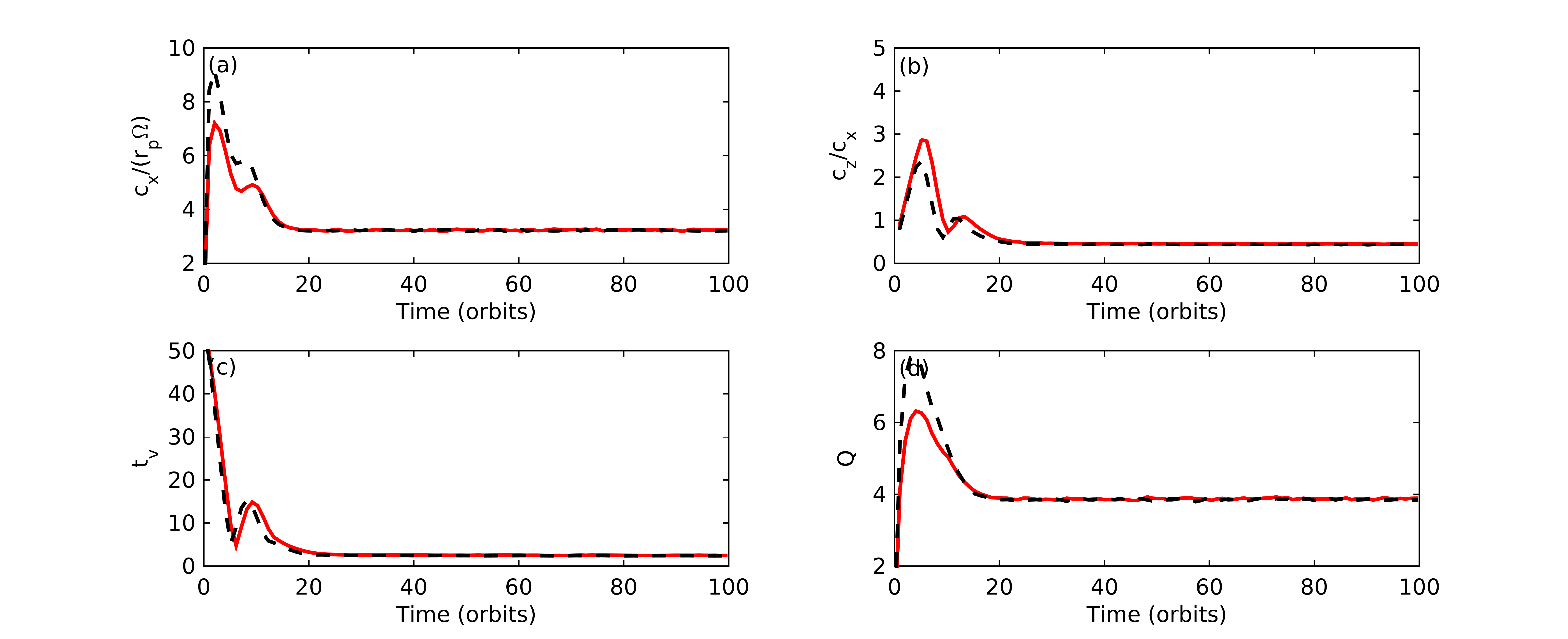}
\caption{Comparison of results from our code (solid lines) with those (dashed lines) of reported by \citet{salo1995a}. The four plots show variation with time (in number of orbits around Saturn) of (a) velocity dispersion $c_x$ in $x$-direction, (b) the ratio $c_z/c_x$, (c) the ring's effective vertical thickness $t_v$ (in particle radii), and (d) the Toomre parameter $Q$.}
\label{fig:code_val1}
\end{figure}

Second, we compared results of our simulations with thousands of particles with existing studies. We simulated a system with optical depth of $0.1$ consisting of $31,830$ self-gravitating particles of density $900\;$kg m$^{-3}$, equaling that of solid ice,   radius $1\;$m, and coefficient of restitution 0.1. Particles were positioned randomly in a square shaped test-section of side $1000\;$m, situated at a distance of $100,000\;$km from a spherical central body of size and mass equal to Saturn's. Particles were initialized with Keplerian velocities. \Cref{fig:code_val1} shows results are in good agreement  with those from existing studies \citep{salo1995a}.

\par
\begin{figure}[!h]
\centering
\includegraphics[width=\textwidth]{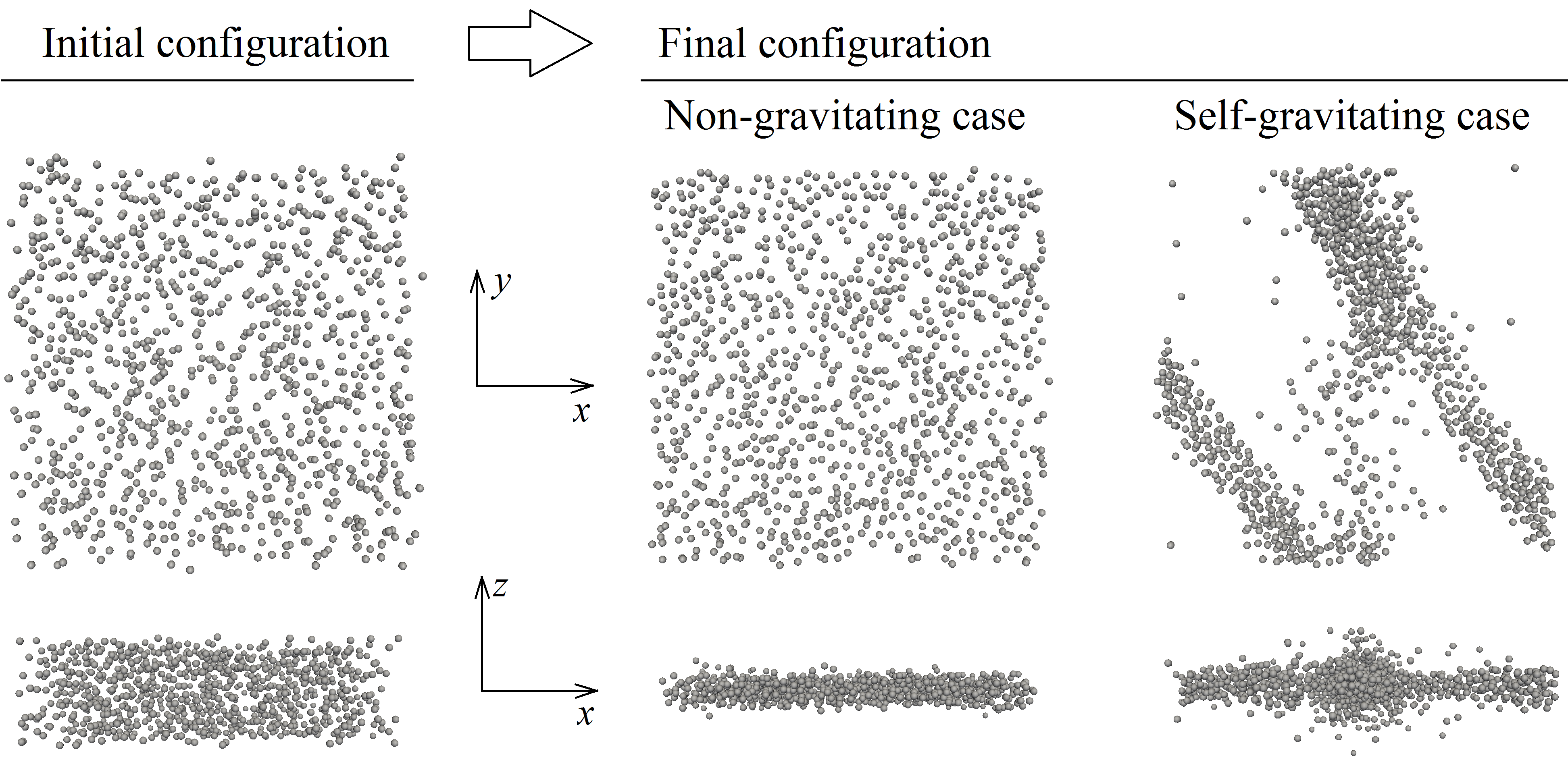}
\caption{Figure shows initial and final snapshots of a local simulation patch viewed along the $z$- and $y$- directions. We employ the same system setup as in Fig.3 of \citet{salo1995a}. There are $1000$ particles of density $900\;$kg m$^{-3}$ and  radius $1\;$m. The TS is square shaped, has size $65\;$m, and is situated at a distance of $100,000\;$km from a spherical central body of dimensions and density equal to Saturn. Initially particles were placed randomly throughout the TS and had Keplerian velocities. The coefficient of restitution $\epsilon_n = 0.5$.}
\label{fig:code_val2}
\end{figure}

 \Cref{fig:code_val2} shows the initial and final snapshots of a non-gravitating and a self-gravitating simulation with the Toomre parameter $Q \sim 0.8$. In the latter case, transient self-gravity wakes \citep{schmidt2009a} are clearly visible, as expected.


\section{Parameters of study} \label{par_study}
\par Here we define parameters employed to characterize the ring dynamics.

\begin{enumerate}
\item \textbf{Velocity dispersion} in a particular direction ($c_x,\;c_y,\;c_z$) corresponds to the root mean square of the velocity fluctuations in that respective direction. The values that we report have been computed `locally'. This means that initially, velocity dispersion is calculated for each particle $k$ taking into account all particles within a representative volume element (RVE) centered on particle $k$, and then averaged over all $N$ particles. This can be defined as
\begin{equation}
	c_j = \frac{\sum_{k=1}^N \sqrt{\langle {(v^{(i)}_j - u_j)}^2 \rangle_k}}{N},
\end{equation}
where $\langle \cdot \rangle_k = \sum_{i=1}^{n_{RVE}} \left(\cdot\right)/n_{RVE}$ is the  average over all particles in a representative volume element (RVE),   $u_j = \langle{v^{(i)}_j}\rangle$ is the mean velocity in the $j^{th}$ direction, and $v^{(i)}_j$ is the effective velocity of $i^{th}$ particle in the $j^{th}$ direction:
\begin{equation*}
	 v^{(i)}_1 = \dot{x}_i,\;v^{(i)}_2= \dot{y}_i - \Psi x_{i}\quad \textrm{and}\quad v^{(i)}_3 = \dot{z}_i.
\end{equation*}
In our simulations the RVE is a sphere of diameter equal to twenty particle radii.

%

\item \textbf{Granular temperature ($T_g$)} is the macroscopic analog of thermodynamic temperature defined in the kinetic theory of gases, which accounts for the fluctuational motion of particles \citep{campbell1990a}. The granular temperature  is defined by
\begin{equation}
	T_g = \frac{1}{3} \sum_{j=1}^3 c_j^2.
\end{equation}
In the above the contribution of fluctuations in the particles' rotational motion is ignored, as our particles are taken to be smooth and non-rotating. 

\item \textbf{Optical depth ($\tau$)} is defined  for a system of identical particles by
\begin{equation}
\tau = \frac{N \pi r_p^2}{x_{TS}\;y_{TS}},
\end{equation}
where $N$ is the total number of particles in the TS that has dimensions $x_{TS}$ and $y_{TS}$.

\item \textbf{Effective vertical thickness ($t_v$)} is the effective vertical thickness of the ring, computed by
\begin{equation}
	{t_v} = \frac{1}{r_p}\sqrt{\frac{12}{N} \sum_i^N z_i^2},
\end{equation}
where $z_i$ is the $z$-coordinate of the $i^{th}$ particle and $r_p$ is the particle radius. 

\item \textbf{Effective radial width ($w_r$)} is the radial extent of the ring, relevant only to T2 simulations, and defined by
\begin{equation}
	{w_r} = \frac{1}{r_p}\sqrt{\frac{12}{N} \sum_i^N x_i^2}.
\end{equation}
where $x_i$ is the $x$-coordinate of the $i^{th}$ particle.

\par Correspondingly, the \textbf{effective spreading rate ($s_{w_r}$)} is the ratio of change in $w_r$ during an averaging period to the total averaging time. We note that the trends in spreading rates are observed to be proportional to the ring's radial width, i.e., simulations with higher radial widths have correspondingly higher spreading rates. This ensures that the reported trends would have held if the simulations were to be continued further.

\item \textbf{Impact frequency ($\omega_c$)} is the total number of impacts taking place in the TS per second per particle. 

\item \textbf{Number density ($n_s$)} is the number of particles per unit volume.

\item {\bf Pressures at $\pmb z\;\textrm{or}\;\pmb x \pmb = \pmb 0$ planes} are, respectively, the force per unit area acting on the upper half of the ring at the plane $z=0$, or the radially-exterior half of the ring separated by the plane $x=0$. The latter is useful only for T2 simulations. This pressure $(p_x, p_x)$ involves momentum transfer due to both particle collisions ($p_{z,coll}$, $p_{x,coll}$) and the (streaming) motion of particles across a plane ($p_{z,str}$, $p_{x,str}$).  

To estimate the collisional contribution only those interactions are considered which happen at or across the $z$ or $x = 0,$ plane. Similarly, the streaming pressure is computed by accounting for the momentum carried by the particles crossing these planes during each time step. Observing the TS during a time-interval $\delta t$, the pressure at $t+\delta t$ may be expressed as follows:
\begin{equation}
	p_{z,coll} = \left( \sum_{z_i(t)\;z_j(t)<0}\;F_{z, coll}^{ij} \right)/A_z,\;\;
	p_{z,str} = \left( \sum_{z_i(t)\;z_i(t+\delta t)<0}\frac{\delta  q^i_{z}}{\delta t} \right)/A_z,\;\;
\textrm{and}\;\;p_z = p_{z,coll}\;+\;p_{z,str},
\end{equation}
where $A_z = w_r y_{TS}$ is the cross-sectional of the $z=0$ plane, $F_{z, coll}^{ij}$ is the collisional force between particles that lie on opposite sides of the $z=0$ plane, and $\delta q^i_{z}$ is $z$-momentum carried by the $i^{th}$ particle into (positive) or away (negative) from the top half of the ring across $z=0$ in time $\delta t$. The summation for $p_{z,coll}$ is over all colliding pairs $\{i,\;j\}$ that have their centers on the opposite sides of the $z=0$ plane at time $t$. The summation for $p_{z,str}$ is over all particles that cross the $z=0$ plane during the interval $\delta t$. The forces acting on the particles are assumed to be constant during time-interval $\delta t$. The final values reported in the main text are obtained after time-averaging the pressures estimated by the above formulae.

Pressure calculations in the $x$ direction are defined similarly. 

\item \textbf{Roche radius ($r_{Roche}$)} refers to the smallest distance from the central body at which two synchronously rotating identical particles can stay in contact with each other due to mutual gravity. For an axisymmetric ellipsoidal central body $r_{Roche}$ is obtained as a solution to
\begin{equation}
{r_{Roche}}^6 + {r_{Roche}}^4 (a_z^2 -a_r^2) - (12\rho'a_r^2 a_z)^2 = 0,\label{eq:rRoche}
\end{equation}
where $\rho' = \rho_{cb}/\rho_{p}$, and $a_r$ and $a_z$ are the central ellipsoid's semi-major axes. 

\item \textbf{Toomre parameter ($Q$)}  is given by
\begin{equation}
Q = \frac{c_x \kappa}{3.36G\sigma},
\end{equation}
where $\sigma$ is the surface density of the ring and $G$ is the universal gravitational constant. The Toomre parameter characterizes the onset of an instability whereby a particle disk with self-gravity is susceptible to growth of axi-symmetric disturbances \citep{toomre1964a}. This instability gets manifested and self-gravity wakes are observed when $Q$ falls below $1$ \citep{schmidt2009a} -- although $N$-body simulations \citep{salo1995a} report this even for $Q\sim 2-3$; see also \Cref{fig:code_val2}. 

\end{enumerate}

\newpage

\section{Trends in the ring properties with variation in the individual characteristic frequencies}\label{trends_char_freq}

\begin{figure} [h!]
\centering
\adjincludegraphics[width=\textwidth,trim={{0.075\width} {0.50\height} {0.075\width} 0},clip]{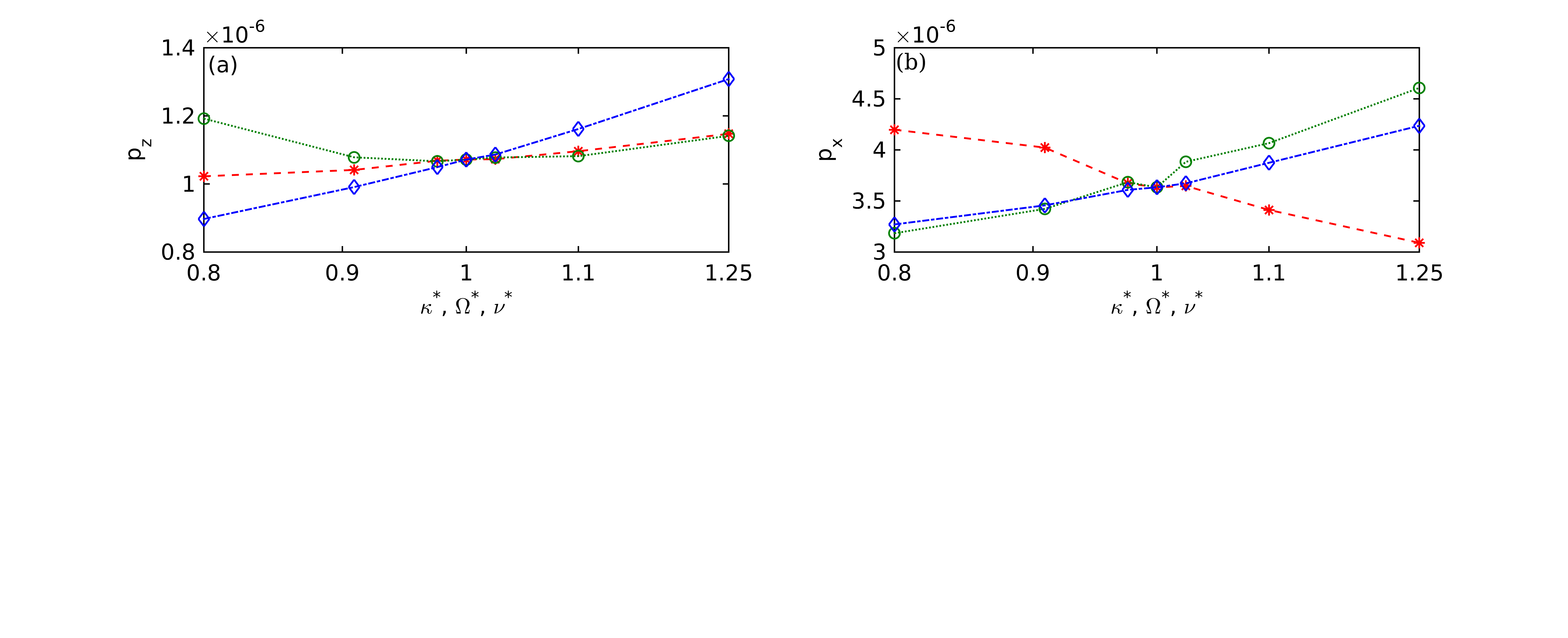} 
\adjincludegraphics[width=\textwidth,trim={{0.075\width} {0.50\height} {0.075\width} 0},clip]{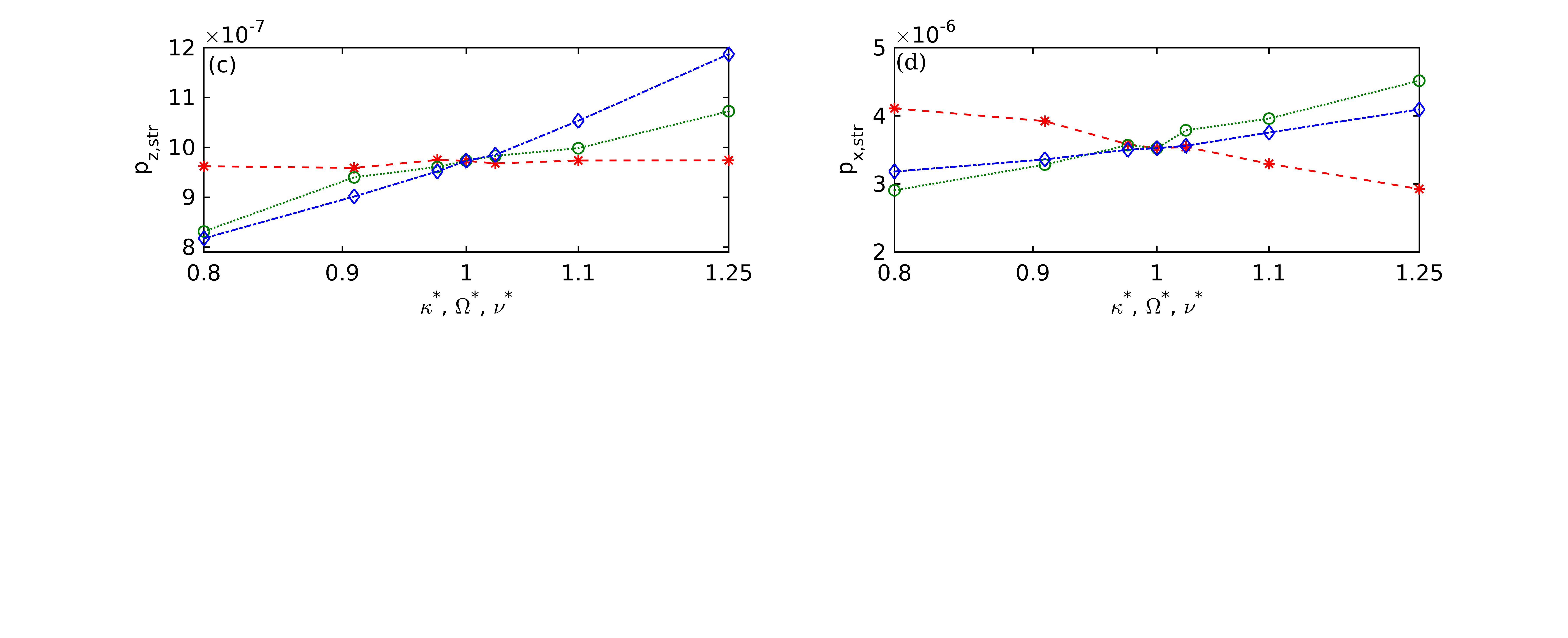}
\adjincludegraphics[width=\textwidth,trim={{0.075\width} {0.50\height} {0.075\width} 0},clip]{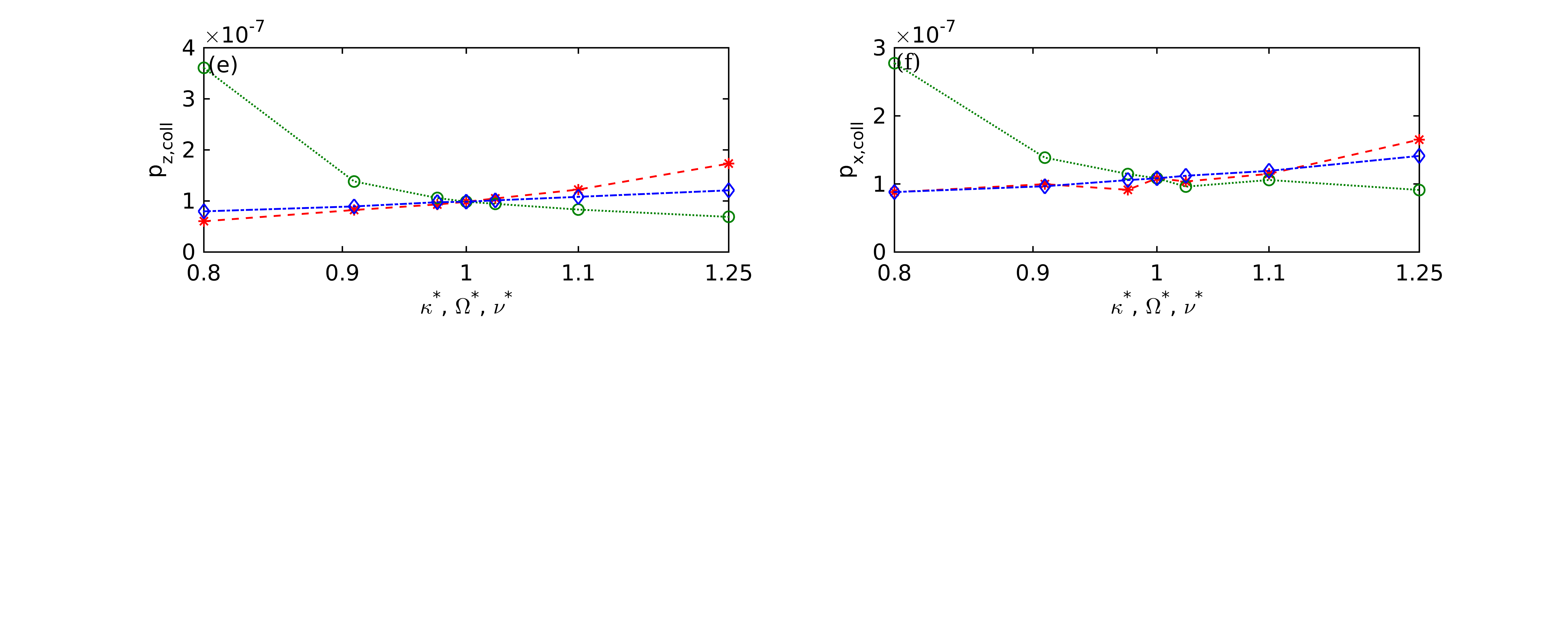}
\caption{Self-gravitating T2 simulations: Variations with characteristic frequencies in the pressure (in N/m$^2$) at $z=0$ ($p_{z}$) and $x=0$ plane ($p_{x}$), and their streaming ($p_{z,str}$, $p_{x,str}$) and collisional ($p_{z,coll}$, $p_{z,coll}$) components. Left and right columns report results for the $z=0$ and $x=0$ plane, respectively. Symbols correspond to  different characteristic frequencies: `{\color{red}- -$\ast$- -}' for $\kappa^*$, `{\color{green}$\cdot\cdot\circ\cdot\cdot$}' for $\Omega^*$ and `{\color{blue}$\cdot$ -$\diamond$- $\cdot$}' for $\nu^*$.}
\label{fig:T2_char_freq_par2}
\end{figure}

\begin{figure} [!h]
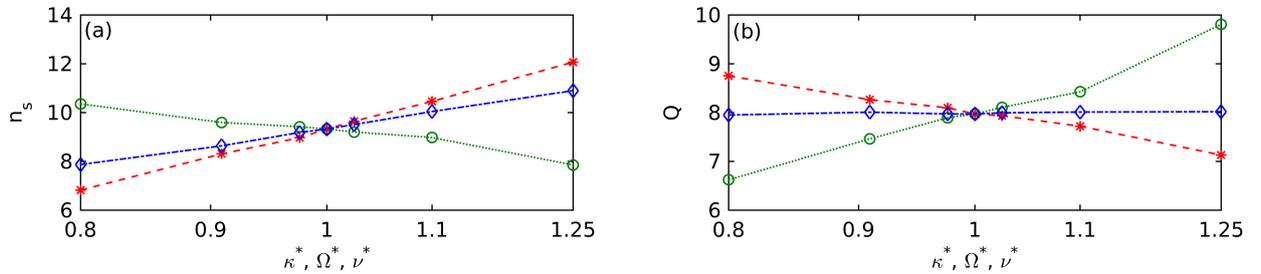

\centering
\adjincludegraphics[width=\textwidth,trim={{0.075\width} {0.00\height} {0.075\width} {0.51\height}},clip]{char_freq_1.png}
\caption{Self-gravitating T2 simulations: Variations with characteristic frequencies in (a) number density ($n_s$, in m$^{-3}$) and (b) Toomre parameter ($Q$). Symbols correspond to the different characteristic frequencies: `{\color{red}- -$\ast$- -}' for $\kappa^*$, `{\color{green}$\cdot\cdot\circ\cdot\cdot$}' for $\Omega^*$ and `{\color{blue}$\cdot$ -$\diamond$- $\cdot$}' for $\nu^*$.}
\label{fig:T2_char_freq_par4_1}
\end{figure}

\begin{figure} [h!]
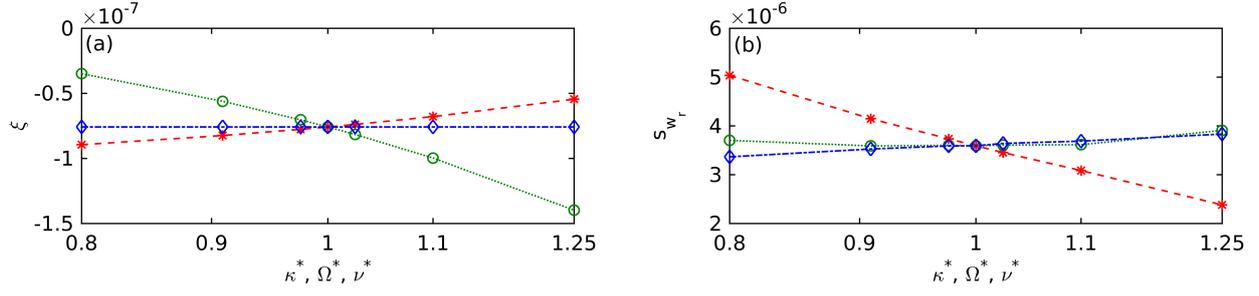

\centering
\adjincludegraphics[width=\textwidth,trim={{0.075\width} {0.0\height} {0.075\width} {0.505\height}},clip]{char_freq_2.png} 
\caption{Self-gravitating T2 simulations: Variations with characteristic frequencies in (a) radial tidal force ($\xi$, in s$^{-2}$) and (b) effective radial spreading rate ($s_{w_r}$, in particle radii/s). Symbols correspond to the different characteristic frequencies: `{\color{red}- -$\ast$- -}' for $\kappa^*$, `{\color{green}$\cdot\cdot\circ\cdot\cdot$}' for $\Omega^*$ and `{\color{blue}$\cdot$ -$\diamond$- $\cdot$}' for $\nu^*$.}
\label{fig:T2_char_freq_par4_2}
\end{figure}

\Cref{fig:T2_char_freq_par2,fig:T2_char_freq_par4_1,fig:T2_char_freq_par4_2} show variations of several ring properties with characteristic frequencies $\kappa^*, \Omega^*$ and $\nu^*$. The trends observed are summarized in \Cref{table:char_freq_var}.

\LTcapwidth=\textwidth
\begin{longtable}{ m{0.08\textwidth}  m{0.03\textwidth}  m{0.03\textwidth}  m{0.03\textwidth}  m{0.05\textwidth}  m{0.08\textwidth}  m{0.03\textwidth}  m{0.03\textwidth}  m{0.03\textwidth}  m{0.05\textwidth}  m{0.08\textwidth}  m{0.03\textwidth}  m{0.03\textwidth}  m{0.03\textwidth} } 

 \hline
 Property & $\kappa^*$ & $\Omega^*$ & $\nu^*$ && Property & $\kappa^*$ & $\Omega^*$ & $\nu^*$ && Property & $\kappa^*$ & $\Omega^*$ & $\nu^*$\\ 
\hline 
\endfirsthead

\hline
 Property & $\kappa^*$ & $\Omega^*$ & $\nu^*$ && Property & $\kappa^*$ & $\Omega^*$ & $\nu^*$ && Property & $\kappa^*$ & $\Omega^*$ & $\nu^*$\\ 
 \hline
\endhead

\hline
\endfoot

\endlastfoot

$p_{z}$&$\uparrow$&$\updownarrow$&$\uparrow\uparrow$&&
$p_{z,str}$ &$\sim$&$\uparrow$&$\uparrow$&&
$p_{z,coll}$ &$\uparrow\uparrow$&$\downarrow\downarrow\downarrow$&$\uparrow$\\

$p_{x}$ &$\downarrow$&$\uparrow$&$\uparrow$&&
$p_{x,str}$ &$\downarrow$&$\uparrow$&$\uparrow$&&
$p_{x,coll}$ &$\uparrow$&$\downarrow\downarrow$&$\uparrow$\\

$T_g$ &$\downarrow\downarrow$&$\uparrow\uparrow$&$\downarrow$&&
$\omega_c$&$\uparrow\uparrow$&$\downarrow\downarrow\downarrow$&$\uparrow$&&
$n_s$ &$\uparrow$&$\downarrow$&$\uparrow$\\

$t_v$ &$\downarrow$&$\uparrow$&$\downarrow$&&
$w_r$ &$\downarrow\downarrow$&$\updownarrow$&$\uparrow$&&
$s_{w_r}$ &$\downarrow\downarrow$&$\updownarrow$&$\uparrow$\\

$Q$ &$\downarrow\downarrow$&$\uparrow\uparrow$&$\uparrow$&&
$\xi$ &$\downarrow$&$\uparrow\uparrow$&$\sim$&&&&&\\

$c_x$ &$\downarrow\downarrow$&$\uparrow\uparrow$&$\downarrow$&&
$c_y$ &$\downarrow$&$\downarrow\downarrow$&$\downarrow$&&
$c_z$ &$\downarrow$&$\uparrow$&$\sim$\\

 \hline
\multicolumn{3}{l}{} \\[-10pt] 
 \caption{Summary of trends observed in \Cref{fig:T2_char_freq_par1,fig:T2_char_freq_par2,fig:T2_char_freq_par4_1,fig:T2_char_freq_par4_2}. A proportional increase in a property with the corresponding characteristic frequency is indicated by `$\uparrow$',  `$\downarrow$' implies the inverse behavior, `$\updownarrow$' corresponds to non-monotonic variation and `$\sim$' indicates no or negligible change. Number of arrow signs signify the relative magnitude of the rate of change.}
\label{table:char_freq_var}
\end{longtable}

\end{document}